\documentclass[iop,apj,revtex4,twocolappendix]{emulateapj} 

\usepackage{longtable}
\usepackage{lscape}
\usepackage{natbib}
\usepackage{psfig}
\usepackage{graphicx}
\usepackage{color}
\usepackage{times}

\def\farcsn{\hbox{$\ \!\!^{\prime\prime}$}}

\def\kms{$\mathrm{km\;s}^{-1}$}
\def\kmsmpc{$\mathrm{km\;s^{-1}\;Mpc^{-1}}$}

\def\lsunpcsq{L$_{\odot} \rm pc^{-2}$}

\def\msun{\ensuremath{M_{\odot}}}
\def\msunh{\ensuremath{h^{-1} M_{\odot}}}

\def\mass{mag arcsec$^{-2}$}

\def\re{\ensuremath{R_{\rm e}}}

\def\sigmae{$\sigma_{\rm e}$}
\def\mstar{\ensuremath{M_{\star}}}

\def\mvir{\ensuremath{M_{\rm dyn}}}

\def\xmm{XMM2235}
\def\xcs{XMM2215}
\def\cl{Cl0332}
\def\hst{{\it HST\/}}
\shorttitle{The Fundamental Plane of cluster galaxies at $1.4<z<1.6$} 
\shortauthors{Beifiori et al.}

\begin{document}  
  

\title{The KMOS Cluster Survey (KCS) I: The fundamental plane and the formation ages of cluster galaxies at redshift
  $1.4<z<1.6$\footnotemark[1]}

\footnotetext[1]{Based on observations obtained at the Very Large
  Telescope (VLT) of the European Southern Observatory (ESO), Paranal,
  Chile (ESO program IDs: 092.A-0210; 093.A-0051; 094.A-0578;
  095.A-0137(A); 096.A-0189(A); 097.A-0332(A)). This work is further
  based on observations taken by the CANDELS Multi-Cycle Treasury
  Program with the NASA/ESA HST, which is operated by the Association
  of Universities for Research in Astronomy, Inc., under NASA contract
  NAS 5-26555.}


\author{Alessandra Beifiori\altaffilmark{1,2}, J. Trevor
  Mendel\altaffilmark{2,1}, Jeffrey C.~C. Chan\altaffilmark{1,2,3},
  Roberto P. Saglia\altaffilmark{2,1}, Ralf Bender\altaffilmark{1,2},
  Michele Cappellari\altaffilmark{4}, Roger L. Davies\altaffilmark{4},
  Audrey Galametz\altaffilmark{2,1}, Ryan
  C. W. Houghton\altaffilmark{4}, Laura J. Prichard\altaffilmark{4},
  Russel Smith\altaffilmark{5}, John P. Stott\altaffilmark{4,6}, David
  J. Wilman\altaffilmark{1,2} , Ian J. Lewis\altaffilmark{4}, Ray
  Sharples\altaffilmark{5}, and Michael Wegner\altaffilmark{1}}
\affil{\altaffilmark{1}{Universit\"{a}ts-Sternwarte M\"{u}nchen, Scheinerstrasse 1,
  D-81679 M\"{u}nchen, Germany; \email{beifiori@mpe.mpg.de}}
\altaffilmark{2}{Max-Planck-Institut f\"{u}r Extraterrestrische Physik,
  Giessenbachstra\ss e 1, D-85748 Garching, Germany}
\altaffilmark{3}{Department of Physics and Astronomy, University of California, Riverside, CA 92521, USA}
\altaffilmark{4}{Sub-department of Astrophysics, Department of
  Physics, University of Oxford, Denys Wilkinson Building, Keble Road,
  Oxford OX1 3RH, UK} 
\altaffilmark{5}{Centre for Advanced Instrumentation, Department of
  Physics, Durham University, South Road, Durham DH1 3LE, UK}
\altaffilmark{6}{Department of Physics, Lancaster University, Lancaster LA1 4YB, UK}
}

\begin{abstract}
 
  We present the analysis of the fundamental plane (FP) for a sample
  of 19 massive red-sequence galaxies (\mstar$>4\times10^{10}$\msun)
  in 3 known overdensities at $1.39<z<1.61$ from the KMOS Cluster
  Survey, a guaranteed time program with spectroscopy from the K-band
  Multi-Object Spectrograph (KMOS) at the VLT and imaging from the
  {\it Hubble Space Telescope}.  As expected, we find that the FP
  zero-point in $B$ band evolves with redshift, from the value 0.443
  of Coma to $-0.10\pm0.09$, $-0.19\pm0.05$, $-0.29\pm0.12$ for our
  clusters at $z=1.39$, $z=1.46$, and $z=1.61$, respectively.  For the
  most massive galaxies ($\log M_{\star}/M_{\odot}>11$) in our sample,
  we translate the FP zero-point evolution into a mass-to-light-ratio
  $M/L$ evolution finding $\Delta \log M/L_{B}=(-0.46\pm0.10)z$,
  $\Delta \log M/L_{B}=(-0.52\pm0.07)z$, to
  $\Delta \log M/L_{B}=(-0.55\pm0.10)z$, {   respectively}.  We
  assess the potential contribution of the galaxies structural {  
    and stellar velocity dispersion} evolution to the evolution of the
  FP zero-point and find it to be {   $\sim$6-35\%} {   of the FP
    zero-point evolution}.  The rate of $M/L$ evolution is consistent
  with galaxies evolving passively.  {   By using single stellar
    population models, we find an average age of
    $2.33^{+0.86}_{-0.51}$ Gyr for the $\log M_{\star}/M_{\odot}>11$
    galaxies in our massive and virialized cluster at $z=1.39$,
    $1.59^{+1.40}_{-0.62}$ Gyr in a massive but not virialized cluster
    at $z=1.46$, and $1.20^{+1.03}_{-0.47}$ Gyr in a protocluster at
    $z=1.61$. After accounting for the difference in the age of the
    Universe between redshifts, the ages of the galaxies in the three
    overdensities are consistent within the errors, with possibly a
    weak suggestion that galaxies in the most evolved structure are
    older.}

\end{abstract}


%galaxies: clusters: individual (name) -- 

%galaxies: fundamental  parameters --
%\date{{\it Draft version on \today}}  

\section{Introduction} 
\label{intro}

In the local universe early-type galaxies lie along a tight relation, the
``fundamental plane'' \citep[FP,
e.g.,][]{Djorgovski1987,Dressler1987}, connecting their surface
brightness within the effective radius $\langle I_{\rm e}\rangle$,
effective radius \re, and velocity dispersion within the effective
radius $\sigma_{\rm e}$.  The FP is tilted with respect to the virial
prediction; the tilt is related to the change of the mass-to-light
($M/L$) ratio with galaxy luminosity, due to a contribution of
variations of the galaxies stellar populations, dark matter fractions and
non-homology
\citep[e.g.,][]{Bender1992,Renzini1993,Jorgensen1996,Renzini2006,Cappellari2006,Cappellari2013a,Scott2015,Cappellari2016}.  While there is still
debate on whether the coefficients of the FP remain constant up to
$z\sim1$ (see \citealt{Holden2010}, \citealt{Saglia2010} and
\citealt{Jorgensen2013}), there is a clear consensus about the
variation of its zero-point with redshift. The zero-point can vary as
a result of evolving $M/L$ \citep{Faber1987} caused by the change in
galaxy luminosity due to the younger stellar population at
high-$z$
\citep[e.g.,][]{vanDokkum1996,Bender1998,Kelson2000,Gebhardt2003,Wuyts2004,Holden2005,Jorgensen2006,vanDokkum2007,Holden2010,Toft2012,Jorgensen2014,Bezanson2013b};
some contribution is also expected from the galaxies structural evolution with
redshift \citep[e.g.,][]{Saglia2010,Saglia2016}.

Several papers have shown that intermediate and high-redshift passive galaxies have
smaller sizes \citep[e.g.,][]{Trujillo2007,Newman2012,
  Houghton2012,vanderWel2014, Beifiori2014, Chan2016} and higher
stellar velocity dispersions (e.g., \citealt{Cappellari2009}; \citealt{Cenarro2009};
\citealt{vanDokkum2009};
 \citealt{vandeSande2013}; \citealt{Belli2014a}; \citealt{Belli2014b};
 \citealt{Belli2015}) compared to
their local counterparts of the same mass or fixed cumulative number
density \citep[e.g.,][]{Brammer2011,
  Papovich2011,Patel2013,vanDokkum2013,Muzzin2013}.

Several authors also suggested that environmental effects may accelerate
the size evolution in clusters compared to the field at $z>1.4$, finding
that galaxies in clusters are larger compared to the field galaxies at
the same redshift (e.g., \citealt{Delaye2014},
\citealt{Lani2013}, \citealt{Strazzullo2013}, Chan et al in sub., but see also
\citealt{Saracco2014} and \citealt{Newman2014} for different results).
In the local universe there are instead negligible differences
  between the mean galaxy sizes in
  different environments \citep[e.g.,][]{Cappellari2013c,Huertas_Company2013}.

The rate of the $M/L$ evolution is described by $\Delta \log M/L
\propto z$, as seen from both samples of massive cluster
\citep[e.g.,][]{vanDokkum1996,Jorgensen2006,Barr2006,Holden2010,Saglia2010} and field
galaxies
\citep[e.g.,][]{vanderWel2005,Treu2005,Saglia2010,vandeSande2014}.
By fitting passively-evolving simple stellar population models, the
$M/L$ evolution can be translated into a formation redshift
corresponding to the epoch of the last major star-formation episode
\citep[e.g.,][]{Tinsley1976}. This technique assumes a
  uniform population across all the galaxies, that the single stellar
  population model approximation holds, and that high-redshift clusters
evolve into a reference low-redshift cluster without any additional
star formation, merging or quenching of star formation.

Following this method, some authors  found that
the stellar populations in galaxies in clusters at $z\leq1$ are older
than in field galaxies \citep[e.g,][]{Gebhardt2003,diSeregoAlighieri2006,vanDokkum2007,Saglia2010},
suggesting an accelerated evolution of passive galaxies in dense
environments.  An age difference between galaxies in clusters and
field could be expected, considering that clusters are formed in the highest
density regions of the Universe, which collapse first.
On the contrary, other authors \citep[e.g.,][]{Treu2005,vanderWel2005,Renzini2006} found that
the stellar mass, rather than environment, is the best predictor of
galaxy ages, with massive galaxies being older.
Further work on the star-formation history of massive galaxies from
the $\alpha/{\rm Fe}$ abundance in the local universe suggested that
the star formation timescales are short and that more massive galaxies
are older compared to those at lower mass, with age differences
between clusters and field galaxies in some case
\citep[e.g.,][]{Thomas2005} and similar ages in other cases
\citep[e.g.,][]{Thomas2010}. The latter is in agreement with the lack of significant
difference in $M/L$ as measured from detailed dynamical models
\citep{Cappellari2006,Cappellari2013a}.

The advent of new IR spectrographs at the VLT and Keck recently
enabled the observation of rest-frame optical spectra for an
increasingly larger number of passive galaxies in the field at $z>1.3$
(e.g., \citealt{vandeSande2011}; \citealt{vandeSande2013};
\citealt{Mendel2015}; Mendel et al in prep. at VLT, and
\citealt{Belli2015,Belli2017} at Keck), some of which were also used
to constrain the galaxies' formation redshift using the FP
\citep[e.g.,][]{vandeSande2014}. At $z>1.3$ differences in the
kinematics and formation ages of passive galaxies in different
environments are currently almost unexplored.  At those redshift,
  we expect to have more constraints on age differences than
  in the local universe, where a 2 Gyr difference in a population of
  $\sim8-10$ Gyr would be challenging to constrain.

  In this paper we investigate the evolution of the FP of massive and
  passive galaxies in dense environments at redshift $1.39<z<1.61$ as
  a part of the KMOS Cluster Survey (KCS, \citealt{Davies2015},
  Davies, Bender et al, in prep). KCS is a guaranteed time observation
  (GTO) program mapping the red sequence of cluster galaxies at
  $1.39<z<1.8$ with the K-band Multi-Object Spectrograph (KMOS,
  \citealt{Sharples2012}, \citealt{Sharples2014}) at the ESO Very
  Large Telescope. The multiplexing and near-infrared capabilities of
  KMOS allow us to simultaneously observe $\geq 20$ galaxies per
  overdensity, and map the rest-frame optical absorption features
  commonly studied in the local Universe
  \citep[e.g.][]{Bender1990,Bender1994} using the KMOS $YJ$ band with
  a resolution of $R\sim3500$.
The combination of our KMOS data with the available {\it Hubble Space
  Telescope} (\hst) imaging allow us to trace the evolution of the FP
of quiescent galaxies in dense environments at $z\geq 1.39$ with one of the
largest samples to date.

The paper is organized as follows. The KCS survey, the cluster and
galaxy sample, \hst\ imaging and KMOS spectroscopic data are presented
in Section~\ref{sec:data}. Measurements of velocity dispersions and
structural parameters are described in Section~\ref{sec:gal_prop}. The
local and intermediate-redshift samples used as reference are
described in Section~\ref{sec:local}. The results are presented in
Section~\ref{sec:FP_clusters} and discussed in
Section~\ref{sec:discussion}. The paper concludes with
Section~\ref{sec:conclusions}. Additional information on the
derivation of the kinematics, the selection functions for our sample,
and the effect of different stellar population models and metallicity
assumptions in our analysis are provided in
Appendices~\ref{sec:sigma_test}, \ref{sec:FP_m_l_only_selection}, and
\ref{subsec:ssp_Z}.

Throughout the paper we assume a standard cosmology with $H_0=70$
\kmsmpc, $\Omega_{\rm m}=0.3$, and $\Omega_{\Lambda}=0.7$. For this
cosmology, 1\farcsn\ corresponds to 8.43 kpc, 8.45 kpc, and 8.47 kpc,
at the mean redshift of our overdensities (XMMU J2235.3-2557 at
$z=1.39$, XMMXCS  J2215.9-1738 at $z=1.46$, and Cl 0332-2742 at
$z=1.61$), respectively.  All magnitudes are in the AB photometric
system \citep{Oke1983}.
Throughout the paper we will use the term "overdensity'' and
"cluster'' interchangeably (but we will highlight when the distinction
is astrophysically meaningful).

\section[]{Sample and Data}
\label{sec:data}

\subsection[]{The KMOS Cluster Survey}
\label{subsec:KCS_sample}

KCS is a 30-night KMOS GTO program performing deep absorption-line
spectroscopy in {\it four main overdensities} at $1.39<z<1.8$ and {\it
  one lower-priority overdensity} at $z=1.04$ to bridge our
high-redshift observations with the local sample.  The sample includes
RCS 234526-3632.6 at $z=1.04$ \citep[][hereafter RCS2345]{Meyers2012},
XMMU J2235.3-2557 at $z=1.39$ \citep[][hereafter \xmm]{Mullis2005,
  Rosati2009}, XMMXCS  J2215.9-1738 at $z=1.46$ \citep[][hereafter
\xcs]{Stanford2006,Hilton2007,Hilton2009,Hilton2010}, Cl 0332-2742 at
$z=1.61$ \citep[][hereafter \cl]{Castellano2007,Kurk2009}, and JKCS 041
at $z = 1.8$ \citep[][]{Newman2014,Andreon2014}.
We observed $\geq 20$ galaxies in the field of each overdensity with the aim of studying the
evolution of kinematics and stellar populations in dense environments
at high redshift. 

The overdensities were selected to have a significant amount of
archival data, spanning from multi-band \hst\ photometry to deep
ground-based imaging, and a large number of
spectroscopically-confirmed members to maximize the galaxy selection
efficiency, concentrating on objects with lower contamination from strong sky
emission or telluric absorption.  

The KMOS patrol field of $7^{\prime}.2$ diameter covers the extent of
the core of our overdensities on the sky ($\sim 3$ Mpc at the redshift
of our overdensities).  The integral-field unit (IFU) dimensions,
$2\farcs8\times2\farcs8$ ($\sim 24 \times 24$ kpc at the redshift of
our overdensities), are generally larger than the size of passive
galaxies at $z\sim1.5$ in most of the cases, allowing us to recover
their total flux within an IFU. Each IFU has 0\farcs2$\times$0\farcs2
spatial pixels.

During the ESO periods P92-P97, KCS targeted a total of 106 galaxies
from our main cluster sample at $1.39<z<1.8$, including 20$-$40
galaxies in each structure. Of those, 67 galaxies were red-sequence
selected, and were observed with exposure times of $\sim 15-20$ hours
on source and seeing $<$1\farcsn. 
This represents one of the largest samples of passive galaxies
homogeneously observed and measured in {\it dense environments} at
$z>1.3$.
During the same ESO periods we also observed $\sim$20 galaxies part
of our lower-priority target RCS2345 with exposure time of
$\sim 9$ hours on source.

%*****************************************************************************
%                FIGURE 1
%*****************************************************************************
\begin{figure}
\centering
\includegraphics[width=\columnwidth]{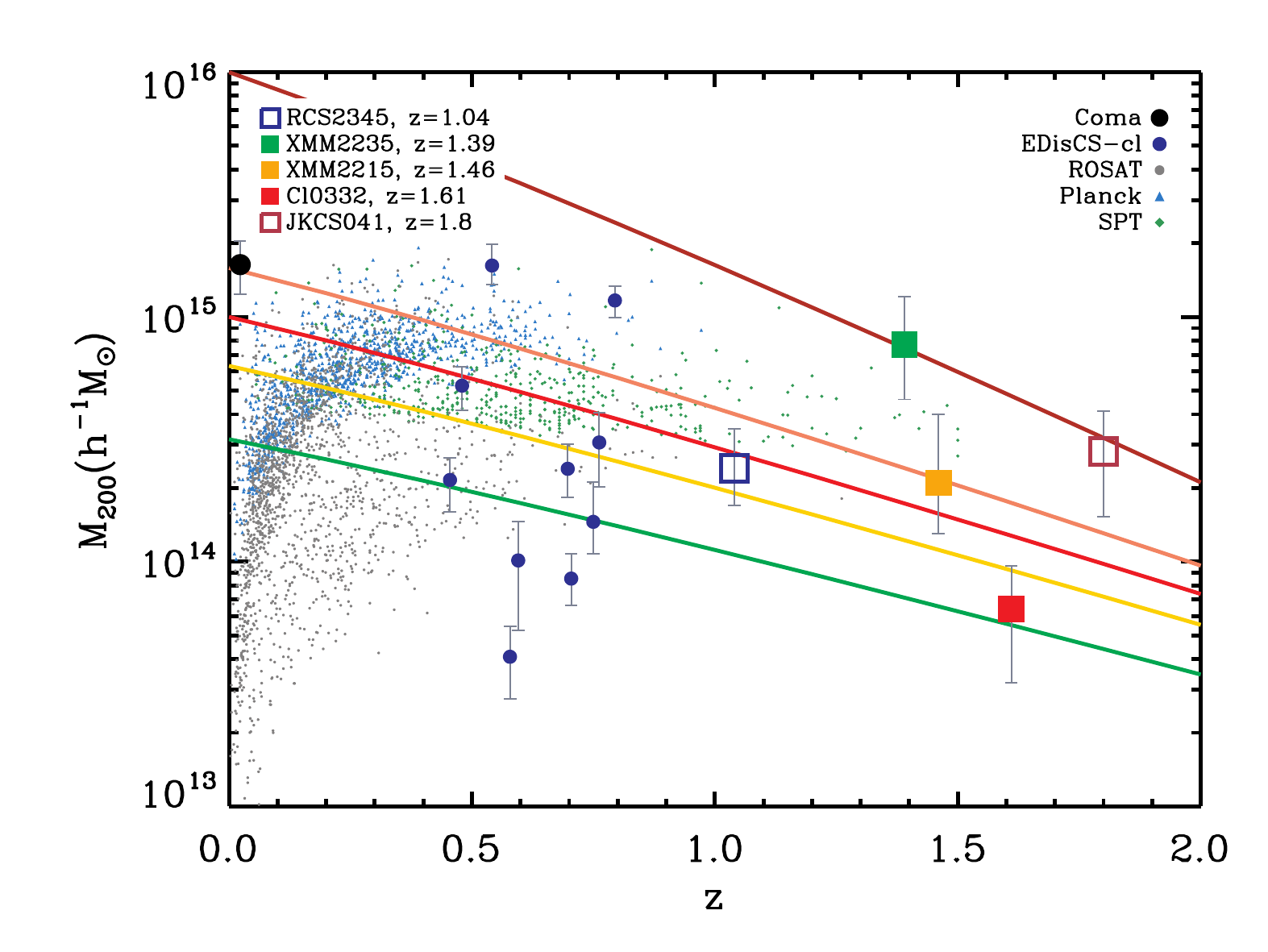}
\caption[] {Cluster mass ($M_{200}$) vs redshift for our full sample
  of KCS clusters from $1.04<z<1.8$.  The green, orange and red filled
  squares are the clusters described in this paper, \xmm, \xcs, and
  \cl. Cluster masses were derived from X-ray data or from the cluster
  velocity dispersion, as tabulated in \citet{Stott2010} and
  \citet{Kurk2009}, respectively. The blue and brown open squares show
  the two additional overdensities from our survey, which will be
  discussed in subsequent papers, RCS2345 at $z=1.04$ in blue ($M_{200}$ from the
  weak-lensing analysis of \citealt{Jee2011}) and JKCS 041 at $z = 1.8$
  in brown ($M_{200}$ from X-ray analysis of \citealt{Andreon2014}).
  We also show the $M_{200}$ for our local reference cluster Coma
  \citep{Lokas2003} and for the subsample of EDisCS clusters and
  groups of \citet{Saglia2010}, which represents our intermediate
  redshift reference (see Section~\ref{sec:local} for
  details). Clusters found through wide-angle surveys, such as ROSAT
  \citep{Piffaretti2011}, {\it Planck} \citep{Planck2016}, and SPT
  \citep{Bleem2015} are also shown.  As comparison we plot the mass
  accretion history of halos of different initial masses $M_{\rm in}$
  derived with the COncentration-Mass relation and Mass Accretion
  History code \citep[COMMAH;][]{Correa2015a,Correa2015b,Correa2015c}
  with continuous lines ($\log M_{\rm in}/M_\odot=16$ in brown, $\log
  M_{\rm in}/M_\odot=15.2$ in pink, $\log M_{\rm in}/M_\odot=15$ in
  red, $\log M_{\rm in}/M_\odot=14.8$ in yellow, $\log M_{\rm
    in}/M_\odot=14.5$ in green).}
\label{fig:cluster}
\end{figure}
%*****************************************************************************

\subsection[]{The sample of overdensities}
\label{subsec:KCS_sample_paper}

In this paper we present the analysis of three overdensities in KCS:
\xmm, \xcs, and \cl, whose general properties we summarize below. 

{\bf \xmm} is a very massive ($M_{\rm 200}=7.7^{+4.4}_{-3.1} \times
10^{14}$\msunh, \citealt{Stott2010}) and virialized cluster
\citep[e.g.,][]{Rosati2009, Stott2010,Jee2011}, which was discovered
by \citet{Mullis2005}. This cluster has a centrally-peaked X-ray
surface brightness profile, suggesting a dynamically relaxed state
\citep{Rosati2009}.  Analysis of stacked spectra
\citep[e.g.,][]{Rosati2009} and colors and scatter of the red
sequence \citep[e.g.,][]{Lidman2008} indicated that massive ($\log
M_\star/M_\odot>11$) galaxies in the core have high formation redshift
($z>3-4$).
Further studies of the luminosity function \citep{Strazzullo2010}
indicate an established high-mass population, suggesting that this
cluster is already at an evolved mass assembly stage.
Its central regions (within $\sim$200 kpc) show no
evidence of star formation \citep[e.g.,][]{Strazzullo2010,Bauer2011},
and, generally, all massive galaxies have low star formation rates
\citep[e.g.,][]{Gruetzbauch2012}. In the outskirts instead, many
galaxies show signatures of star formation as determined from $\rm
H_{\alpha}$ narrow-band imaging
\citep[e.g.,][]{Bauer2011,Gruetzbauch2012}.

{\bf \xcs}, discovered by \citet{Stanford2006}, is a massive
overdensity ($M_{\rm 200}=2.1^{+1.9}_{-0.8} \times 10^{14}$\msunh,
\citealt{Stott2010}), with extended X-ray emission from the hot gas,
suggesting that the cluster is in a relatively advanced evolutionary
stage.  However, the cluster is unlikely to be fully virialized
\citep[e.g.,][]{Ma2015}, as the galaxies velocity distribution is
bimodal \citep[e.g.,][]{Hilton2007,Hilton2010} and there is no clear
brightest cluster galaxy (BCG, see \citealt{Hilton2009},
\citealt{Stott2010}).  The nominal BCG is a
spectroscopically-confirmed member at $\sim$300 kpc from the X-ray
centroid and only marginally brighter than other cluster members.  The
red sequence of \xcs\ is made by relatively faint and low-mass objects
and has a scatter significantly larger than that of local clusters or
\xmm\ \citep[e.g.,][]{Hilton2009}.  From the scatter and the intercept
of the red sequence \citet{Hilton2009} derived a galaxy formation
redshift in the range $z\sim 3-5$.  A significant amount of
red-sequence galaxies show some level of star formation, with
$[\rm O_{II}]$ emission in their observed spectra
\citep[e.g.][]{Hilton2009,Hilton2010} or via narrow band imaging
\citep{Hayashi2010,Hayashi2011,Hayashi2014}.  Some galaxies in the
cluster core show a significant amount of obscured star formation with
substantial emission at 24 $\rm \mu m$ \citep[e.g.][]{Hilton2010} and
in sub-millimeter bands \citep[e.g.,][]{Ma2015, Stach2017}, and there
are also a significant number of AGNs \citep{Hayashi2011}. {
  Moreover, the lack of CO emission in the very center of the
  overdensity also provided some constraints on the possible quenching
  mechanism that galaxies in the cluster experienced
  \citep[e.g.,][]{Hayashi2017}.}

{\bf \cl} was first identified as an overdensity using
photometric redshifts \citep{Castellano2007} and confirmed by the
Galaxy Mass Assembly ultra-deep Spectroscopic Survey (GMASS,
\citealt{Kurk2013}), which targeted photometric redshift-selected
galaxies ($z_{\rm phot}> 1.4$) in the Great Observatories Origins
Survey (GOODS) Southern field \citep{Giavalisco2004}. The mass of this
overdensity ranges from $M_{\rm 200}=6.4^{+0.3}_{-0.3} \times
10^{13}$\msunh\ as measured from the cluster velocity dispersion,
assuming the structure is virialized \citep[][see
Figure~1]{Kurk2009}, to $M_{\rm 200}=1.2\times 10^{14}$\msunh\ as
measured summing the mass of the X-ray groups in the field
\citep[e.g.,][]{Finoguenov2015}.  The members show a bimodal
distribution in velocity, suggesting that the structure is mostly
formed by two main groups with no clear evidence of spatial separation
\citep[e.g.,][]{Kurk2009}.  There is no obvious X-ray emission
throughout the full structure \citep[e.g.,][]{Kurk2009,
  Finoguenov2015}, and the low $S/N$ does not permit a proper
separation from the foreground sources; most of the X-ray emission
comes from the most massive group in the system discussed in
\citet{Tanaka2013}. These findings suggest that \cl\ is likely a
cluster in formation.
Nevertheless, \cl\ shows a clear red sequence in the color-magnitude diagram. The
analysis of the stellar population from stacked spectra shows that
galaxies have a relatively young age, low specific star formation rate
and significant dust extinction \citep{Cimatti2008,Kurk2009}.

Figure~\ref{fig:cluster} shows the cluster masses as a function of
redshift for the KCS sample described in this paper (\xmm, \xcs,
\cl). For reference, we also show the two additional overdensities
that are part of the full KCS sample, JKC S041 and  RCS2345,
and that will be presented in forthcoming papers.

As a local reference we use the Coma cluster (see also
Section~\ref{sec:local}).  Following the prescriptions of
\citet{Hu2003}, we rescale the mass within the virial radius of Coma
from \citet{Lokas2003} to the mass within the radius $\rm R_{200}$ $-$
inside which the average mass density is 200 times the critical
density of the universe $-$ finding $M_{200}=(1.6\pm0.4)\times
10^{15}$\msunh.  For this calculation we used a halo concentration
parameter as described by \citet{Bullock2001}.

Our intermediate redshift reference is a subsample of the EDisCS
clusters and groups (thereafter called ``EDisCS-cl'') used in the
fundamental plane study of \citet{Saglia2010}, for which \hst\ imaging
is available (see also Section~\ref{sec:local}). Their $M_{200}$ was
calculated from their tabulated cluster stellar velocity dispersion
(see Table~4 of \citealt{Saglia2010}) following the prescription of
\citet{Carlberg1997b}.

For comparison, we show public catalogs of clusters from other wide-angle cluster
survey including ROSAT \citep{Piffaretti2011}, {\it Planck}
\citep{Planck2016}, and SPT \citep{Bleem2015}, whose masses $M_{500}$
were converted to $M_{200}$ following the prescriptions of
\citet{Hu2003} with a halo concentration parameter of 5. 

We overplot models for the growth of cluster mass with
time, based on the COncentration-Mass relation and Mass Accretion
History \citep[COMMAH;][]{Correa2015a,Correa2015b,Correa2015c} code,
which uses an analytic model to generate halo mass accretion rates for
a variety of redshifts and cluster masses.

According to those predictions, the two most massive overdensities in
KCS, \xmm\ and JKCS 041, will evolve into clusters with a $M_{200}$
larger than that measured for our local comparison cluster
Coma. Moreover, only two clusters from the ``EDisCS-cl'' sample are
in a mass range similar to the most massive overdensities in KCS.

\subsection[]{Target selection}
\label{subsec:imaging}

\subsubsection[]{Imaging}
\label{subsubsec:imaging}

Archival imaging from the \hst\ Advanced Camera for Survey (ACS) and
the Wide Field Camera 3 (WFC3) are available for our sample (e.g.,
\citealt{Chan2016}; Chan et al., sub.) {  as well as photometric
  data from the ground.}

{  For \xmm\ ACS data are available from program GTO-10698 and
  GO-10496, whereas WFC3 data from program GO/DD-12051;  for \xcs\ ACS data
  come the program GO-10496.}
All the \hst\ data were processed as described by \citet{Chan2016} and
Chan et al. sub.  using {\tt ASTRODRIZZLE} \citep{Gonzaga2012}. For
this paper we used the ACS/$z_{\rm F850lp}$ and WFC3 $Y_{\rm F105w}$,
$H_{\rm F160w}$ bands for the two clusters; see \citet{Chan2016} and
Chan et al. sub.  for more details on the available bands.
For \xcs\ we also used a $J$-band image from the Multi-Object InfraRed
Camera and Spectrograph (MOIRCS) at Subaru Telescope
\citep{Hilton2009} with a seeing FWHM$\sim0$\farcs6.

{  The \hst\ (or MOIRCS) imaging of \xmm\ and \xcs\ have a small
  field of view, which could bias our absolute astrometric
  solutions. Therefore, we applied to our WFC3 and MOIRCS images the
  same astrometry of the $Ks$-band HAWK-I images available for the two
  clusters (\citealt{Lidman2008}, \citealt{Lidman2013}, and C. Lidman,
  private communication)\footnote{Based on data products from
    observations made with ESO Telescopes at the La Silla Paranal
    Observatory under program ID 060.A-9284(H).}, which have a larger
  field of view.}

Source catalogs were produced using SExtractor \citep{Bertin1996},
with the reddest image, or that with the higher resolution used for
detection. We flagged stars with a star-galaxy classification {\tt
  class\_star} $\geq$ 0.9.
We measured both {\tt MAG\_AUTO} and aperture magnitudes within an
aperture of diameter 1\farcsn, which were corrected for Galactic
reddening in the direction of the cluster using the values given by
the the NASA Extragalactic Database extinction law
calculator\footnote{http://ned.ipac.caltech.edu/help/extinction\_law\_calc.html},
which is based on the maps by \citet{Schlafly2011}.
The apertures are larger than the PSF of our images and we expect the
aperture size not to have an impact in the derived magnitudes.
In the final catalog of \xcs, we included only galaxies with
photometric redshift in the range $1.27<z<1.65$ from
\citet{Hilton2009} or with available spectroscopic redshift from the
same paper.  {  The galaxy IDs in Table~\ref{tab:summary_KCS} come
  from our $H_{\rm F160w}$ catalog for \xmm, and from the
  $z_{\rm F850lp}$ catalog for \xcs\ (see also Chan et al, sub).}

For \cl\ we used public ACS and WFC3 mosaics in the $I_{\rm F814w}$,
$J_{\rm F125w}$ bands, respectively, from the Cosmic Assembly Near-IR
Deep Extragalactic Legacy Survey (CANDELS; \citealt{Grogin2011};
\citealt{Koekemoer2011}) as provided by the 3DHST survey
\citep{Skelton2014} as well as the public catalog in the GOODS-S field
\citep{Guo2013,Skelton2014,Momcheva2016}. The magnitudes provided in
these catalogs are total PSF-matched magnitudes. We used those when we
needed total magnitudes, whereas to derive aperture magnitudes we
follow Equation~3 of \citet{Skelton2014} to rescale the provided total
magnitudes to the original 0\farcs7 aperture magnitudes and their
errors.
The use of different apertures between \citet{Skelton2014} and
\citet{Chan2016} does not affect our selection because effective radii
of most galaxies in our sample are smaller than the aperture size.
Magnitudes in \citet{Skelton2014} are extinction corrected using the
maps by \citet{Schlafly2011}.
We selected potential cluster members in the GOODS-S field as those
within $\pm3000$\kms\ of the systemic redshift using the best
spectroscopic, grism or photometric estimation from
3DHST\footnote{With some exceptions for objects with uncertain
  spectroscopic redshifts within our field of view} ({\tt z\_best},
\citealt{Momcheva2016}) and within a region covering the KMOS field of
view (i.e, 7.2 arcmin diameter). This includes both the upper part of
the \citet{Kurk2009} structure and the \citet{Tanaka2013} group.
{  In Chan et al. sub. we derive our own SExtractor catalog of the
  WFC3/$H_{\rm F160w}$ CANDELS images; the IDs we provide in
  Table~\ref{tab:summary_KCS} come from that catalog.}

%*****************************************************************************
%                FIGURE 2
%*****************************************************************************
\begin{figure}
\centering
\includegraphics[width=\columnwidth]{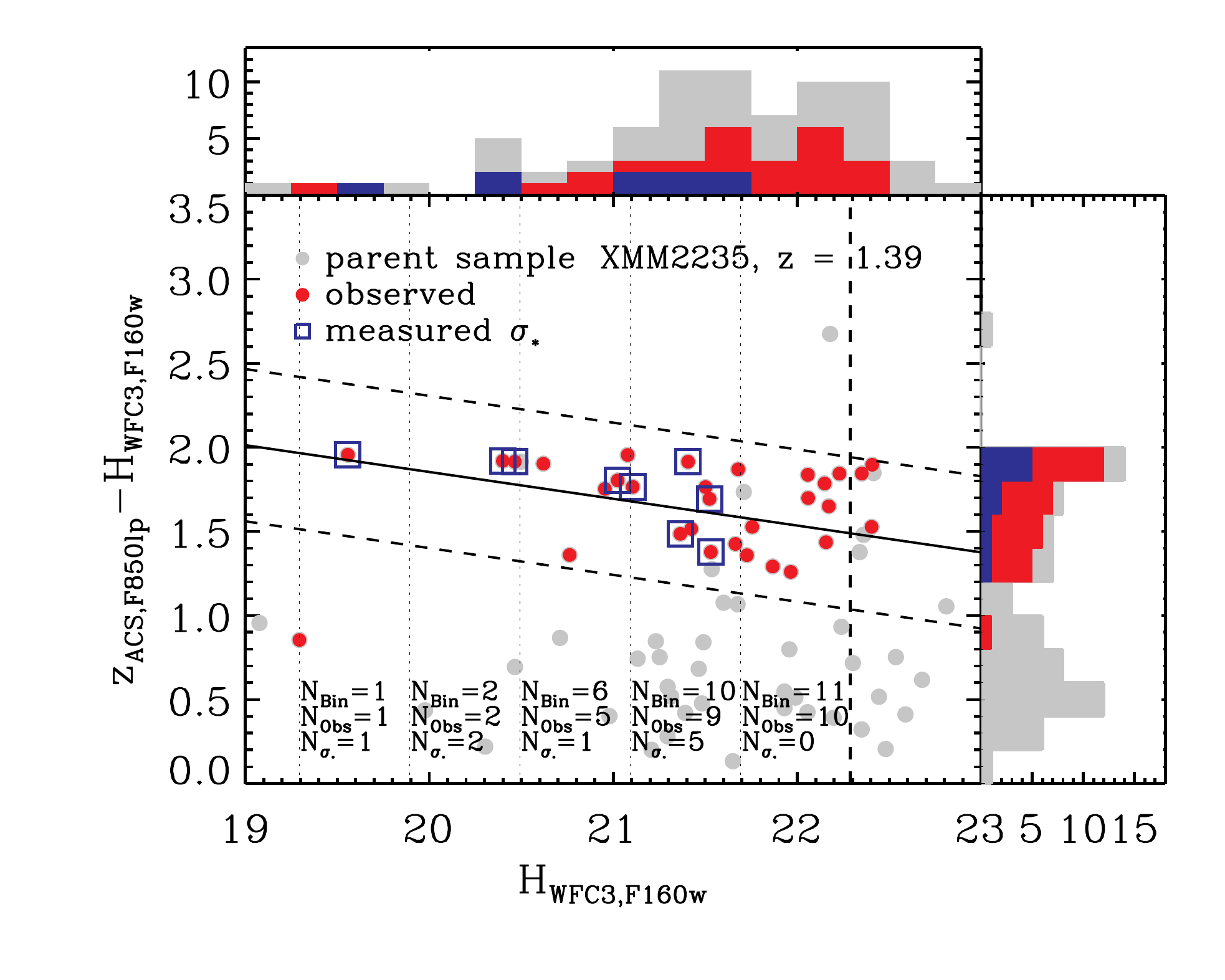}
\includegraphics[width=\columnwidth]{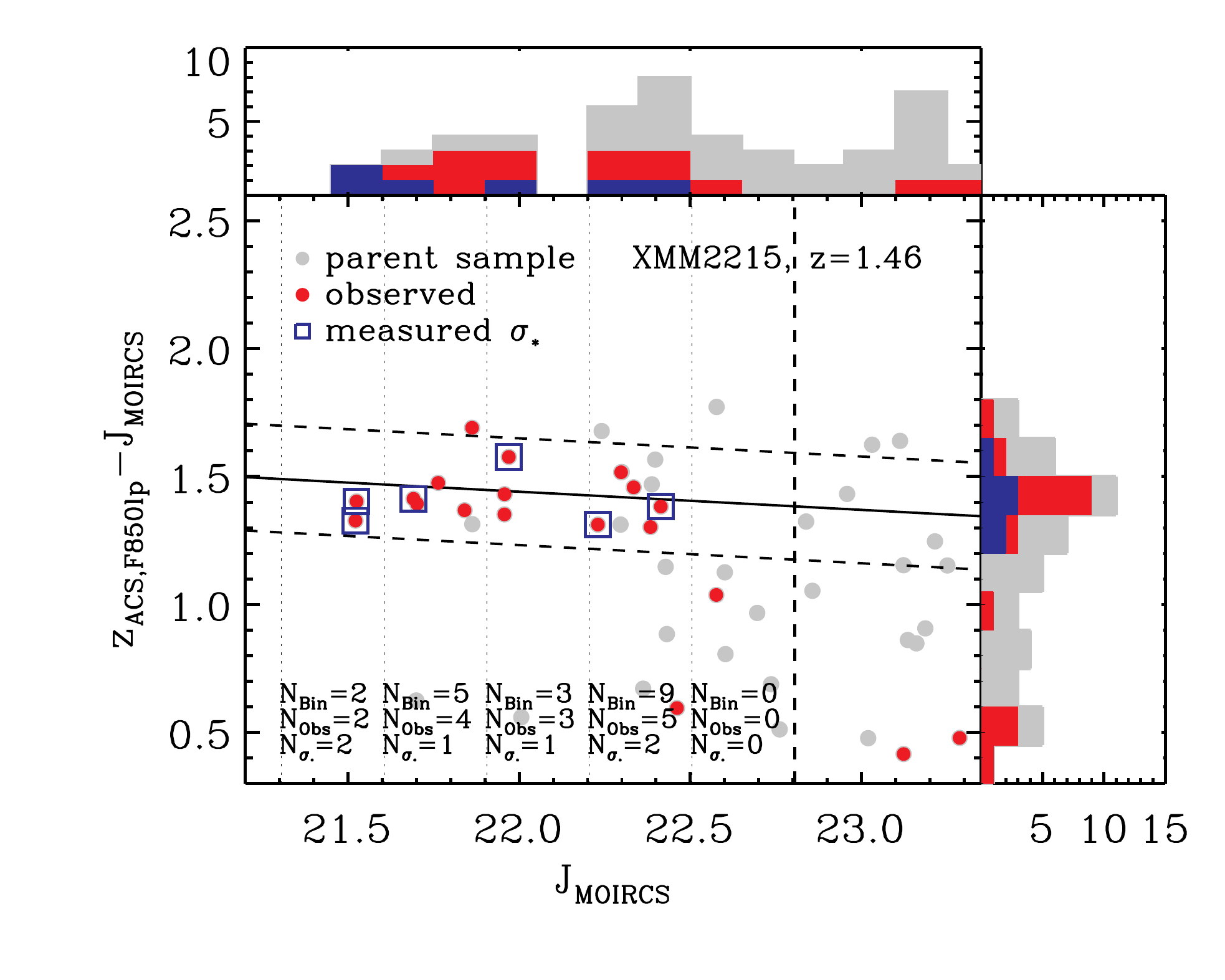}
\includegraphics[width=\columnwidth]{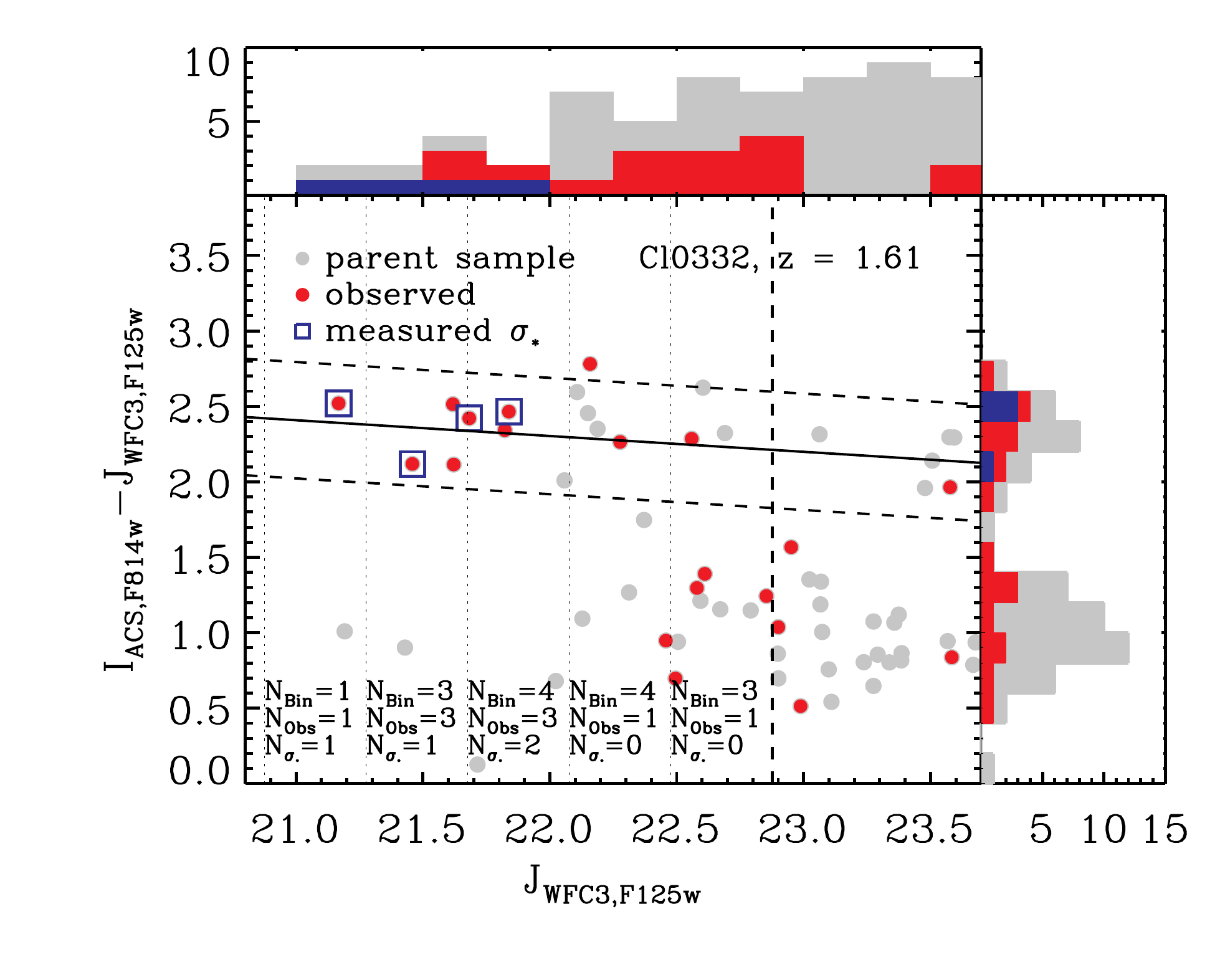}

\caption[] {CMDs showing our parent sample (gray filled circles), the
  galaxies observed by KCS (red filled circles), and the galaxies for
  which we could derive stellar velocity dispersions (blue open
  squares), our dispersion sample, see
  Section~\ref{subsec:sigma}. Black continuous lines are the best-fit
  red sequences and black dashed lines show their $2\sigma$ scatter.
  Vertical dashed lines show the magnitude cut used in our
  completeness analysis (see
  Appendix~\ref{sec:FP_m_l_only_selection}), $J_{\rm F140w}<22.5$,
  rescaled to the band used in the CMD.  Vertical dotted lines show
  the magnitude bins used to evaluate the selection functions for our
  sample (see Appendix~\ref{sec:FP_m_l_only_selection}), with numbers
  of objects per bin $-$ within the red sequence and our magnitude cut
  $-$ shown at the bottom ($\rm N_{\rm Bin}$), as well as objects
  observed with KMOS ($\rm N_{\rm Obs}$) and for which we derived
  stellar velocity dispersion ($\rm N_{\sigma_\star}$).  Histograms
  show the distribution of colors and magnitudes for the three
  samples.}
\label{fig:cmd}
\end{figure}
%*****************************************************************************

\subsubsection[]{Color-magnitude diagrams}
\label{subsubsec:target_sel}

In Fig.~\ref{fig:cmd} we show the resulting color-magnitude diagrams
(CMD) of the three overdensities.  The red-sequence relations are
fitted from the CMDs (solid lines), and their scatter is measured by
marginalizing over the magnitude to obtain the distribution of
galaxies.
 The dashed lines
in the CMDs correspond to the $2\sigma$ scatter derived from a
Gaussian fit of this distribution.  In Table~\ref{tab:summary_KCS} we
give the colors and magnitudes used in the CMDs only for the galaxies
for which we measure stellar velocity dispersion (see
Section~\ref{subsec:sigma}); hereafter we will refer to this as  the
dispersion sample.

Passive galaxies for the KMOS observations were selected to be within
$2\sigma$ from the fitted red sequence of each cluster, to lie within
both the ACS and WFC3 fields of view (when available at the time of
the KMOS observations, see Chan et al sub. for the new WFC3 data
collected after the KMOS observations), and to be bright.
We prioritized bright objects with public spectroscopic redshifts when
available in the red sequence of the CMDs, then objects in the red
sequence with no redshift information, and finally included
lower-priority fillers, either from faint red sequence objects or from
the blue cloud to target emission lines. The latter will be described
in a forthcoming paper (Stott et al, in prep).

In the following, we apply a magnitude limit of $J_{\rm F140W}<22.5$ to
identify bright objects in our analysis, below which only fillers were
included in our allocations.
The actual value of the magnitude limit used in each of our samples
in Fig.~\ref{fig:cmd} was derived applying a color term to
this threshold in $J_{\rm F140W}$ band to match the band used in the
CMD; the color terms were obtained from \citet{Maraston2005} simple
stellar population models with an age $>1$ Gyr.
Based on the simulations described in \citet{Chan2016}, Chan et
al. sub., within our magnitude limit we detect 93\% of the
objects in \xmm, 98\% in \xcs, and 99\% in \cl, respectively.

With a target selection based only on the red sequence, we would
expect a number of interlopers, in particular at faint magnitudes.
The number of possible foreground or background interlopers can be
estimated by comparing the number of objects in the red sequence of
our overdensities with the number of objects we find constructing a
color-magnitude diagram with field data in the same bands of our
overdensities.
We used photometric catalogs from the CANDELS/3DHST deep fields
\citep{Skelton2014,Momcheva2016}, excluding the GOODS-S field because
it is where one of our overdensities reside.
We compared, per bin of magnitude, the
number of objects in the red sequence of our overdensities to the
number of objects of the red sequence found constructing a CMD with
the same band and data from the CANDELS/3DHST deep fields. The
estimated number of interlopers were then rescaled by the ratio
between the area of our observations and the area covered by
CANDELS/3DHST in the bands of our CMDs.
Our estimates are conservative because we assume that there are no
overdensities in the CANDELS/3DHST deep fields.

Based on this comparison, we do not expect to find interlopers in the
bright part of the red sequence ($H_{\rm F160w}<20.5$) for \xmm;
moreover, at these bright magnitudes our selection includes galaxies
with prior spectroscopic redshifts from the literature. At fainter
magnitudes ($21.7<H_{\rm F160w}<22.3$), where we also lack
spectroscopic redshifts, the number of interlopers increases up to 50\% .
For \xcs\ we expect the number of interlopers to vary from $10$\% at
bright magnitudes ($J_{\rm MOIRCS}<21.9$) to $\sim30$\% in the
faintest magnitude bin we targeted ($22.2<J_{\rm MOIRCS}<22.5$).
At the faintest magnitude, the actual effect of the contamination by
interlopers was significantly minimized  by the the additional
information about the photometric or spectroscopic redshift of the
galaxies.
For \cl\ contamination can be more serious, more than 80\% of the brighter 
galaxies  ($J_{\rm F125w}<21.68$) could be interlopers. This effect is
particularly enhanced for this cluster given the lower number of
objects in the red sequence.
The availability of prior information about the galaxy redshifts 
significantly  improved our target selection.

In Chan et al. sub. we use new data collected after the KMOS
observations, and constructed two-color diagrams ($UVR$ and $UVJ$
rest-frame colors) to show that some of the faint objects in \xmm\ and
\xcs\ could potentially either be very dusty star-forming objects
entering in the red sequence or indeed not at the redshift of the
cluster.

%*****************************************************************************
%                Table 1
%*****************************************************************************
\begin{table*}
\begin{tiny}
\begin{center}
\caption{Properties of the galaxies in \xmm,
  \xcs, and \cl.}
\begin{tabular}{l c c c c c c c c c c c c}
\hline
\hline
\noalign{\smallskip}
ID    &       RA       &        DEC    &       Redshift              &
                                                                      $(z-H)$ & $H$& $\log \langle I_{\rm e} \rangle /(L_{\odot}pc^{-2})$      &          \re           &      $n$               &       $q$            & $\sigma_{\rm e} $&$\log M_\star$/\msun      & $\log M_{\rm dyn}$/\msun \\ 
      &      (J2000)          &  (J2000)             &                        &         &  mag         &      &         kpc            &                        &                      &  \kms                   &                         &                                   \\
\hline
352  &   338.836332    &   -25.962342   &  1.3747    & 1.92   &  20.40        &  3.60  $\pm$  0.15     &   1.97 $\pm$   0.34   &   2.47  $\pm$  0.24    &   0.64  $\pm$   0.02  &   223.21   $\pm$  52.92   &  11.23   $\pm$  0.07     &  11.20     $\pm$  0.22     \\
296  &   338.840083    &   -25.957082   &  1.3793    & 1.38   &  21.53        &  3.33  $\pm$  0.19     &   2.39 $\pm$   0.52   &   0.95  $\pm$  0.10    &   0.57  $\pm$   0.02  &   308.40   $\pm$  77.28   &  10.53   $\pm$  0.05     &  11.63     $\pm$  0.24     \\ 
407  &   338.836328    &   -25.960471   &  1.3848    & 1.91   &  20.46        &  3.27  $\pm$  0.30     &   3.94 $\pm$   1.36   &   6.73  $\pm$  1.05    &   0.66  $\pm$   0.03  &   206.73   $\pm$  50.36   &  11.24   $\pm$  0.07     &  11.23     $\pm$  0.26      \\
220  &   338.845102     &  -25.940250   &  1.3902    & 1.69   &  21.52        &  3.37  $\pm$  0.28     &   2.43 $\pm$   0.79   &   5.15  $\pm$  0.75    &   0.75  $\pm$   0.04  &   180.67   $\pm$  50.21   &  10.82   $\pm$  0.06     &  10.98     $\pm$  0.28     \\ 
 36  &   338.829552    &   -25.974256   &  1.3919    & 1.77   &  21.11        &  3.18  $\pm$  0.28     &   3.45 $\pm$   1.11   &   4.17  $\pm$  0.64    &   0.79  $\pm$   0.04  &   163.14   $\pm$  30.85   &  11.04   $\pm$  0.08     &  11.09     $\pm$  0.22     \\
576  &   338.841546    &   -25.949133   &  1.3937    & 1.80   &  21.02        &  3.93  $\pm$  0.08     &   1.41 $\pm$   0.13   &   2.76  $\pm$  0.14    &   0.35  $\pm$   0.01  &   376.71   $\pm$  42.49   &  11.01   $\pm$  0.08     &  11.50     $\pm$  0.11     \\
170  &   338.836838    &   -25.961102   &  1.3949    & 1.96   &  19.56        &  2.55  $\pm$  0.70     &  13.51 $\pm$  10.84   &   3.66  $\pm$  1.22    &   0.62  $\pm$   0.10  &   366.59   $\pm$  43.65   &  11.82   $\pm$  0.07     &  12.41     $\pm$  0.36     \\
433  &   338.829397    &   -25.964279   &  1.3951    & 1.91   &  21.41        &  3.07  $\pm$  0.43     &   3.41 $\pm$   1.70   &   5.61  $\pm$  1.08    &   0.71  $\pm$   0.05  &   232.28   $\pm$  46.95   &  10.95   $\pm$  0.07     &  11.33     $\pm$  0.28     \\
637  &   338.844880    &   -25.951640   &  1.3966    & 1.49   &  21.36        &  3.92  $\pm$  0.08     &   1.35 $\pm$   0.12   &   2.78  $\pm$  0.16    &   0.77  $\pm$   0.01  &   175.32   $\pm$  66.00   &  10.69   $\pm$  0.05     &  10.81     $\pm$  0.33     \\

\hline
\noalign{\smallskip}
ID    &   RA          &      DEC      &            Redshift       & $(z -J)$ &  $J$       &  $\log \langle I_{\rm e} \rangle /(L_{\odot}pc^{-2})$          &   \re                  &      $n$               &       $q$               & $\sigma_{\rm e}$           &$\log M_\star$/\msun    & $\log M_{\rm dyn}$/\msun \\ 
      &                &               &                        &         &  mag         &     &         kpc            &                        &                      &  \kms                   &                         &                                   \\
\hline
 864 &    333.996028  & -17.634061  &   1.4505    &   1.58   &  21.97     &    3.58    $\pm$  0.58     &   1.63 $\pm$  1.09     &    2.01  $\pm$ 0.97    &  0.56  $\pm$     0.09   & 182.56   $\pm$   48.51    &  11.13  $\pm$    0.10      &    10.96  $\pm$  0.37  \\ 
 912 &    333.999518  & -17.633135  &   1.4507    &   1.31   &  22.23     &    4.26    $\pm$  0.20     &   0.73 $\pm$  0.17     &    2.70  $\pm$ 0.48    &  0.62  $\pm$     0.03   & 245.49   $\pm$   77.49    &  10.76  $\pm$    0.08      &    10.84  $\pm$  0.29  \\ 
1006 &    333.984149  & -17.630537  &   1.4559    &   1.40   &  21.52     &    4.13    $\pm$  0.25     &   1.27 $\pm$  0.37     &    4.65  $\pm$ 1.06    &  0.30  $\pm$     0.02   & 296.22   $\pm$   58.11    &  11.11  $\pm$    0.08      &    11.16  $\pm$  0.21  \\ 
 710 &    333.983600  & -17.639072  &   1.4587    &   1.38   &  22.41     &    3.60    $\pm$  0.58     &   1.62 $\pm$  1.09     &    2.92  $\pm$ 1.41    &  0.41  $\pm$     0.07   & 294.19   $\pm$   77.80    &  10.87  $\pm$    0.08      &    11.33  $\pm$  0.37  \\ 
 615 &    334.013234  & -17.641575  &   1.4652    &   1.33   &  21.52     &    3.00    $\pm$  0.18     &   4.84 $\pm$  1.03     &    3.86  $\pm$ 1.89    &  0.60  $\pm$     0.04   & 214.45   $\pm$   65.13    &  11.03  $\pm$    0.12      &    11.49  $\pm$  0.28  \\ 
 781 &    333.993745  & -17.636265  &   1.4703    &   1.41   &  21.69     &    3.34    $\pm$  0.38     &   2.23 $\pm$  0.99     &    0.78  $\pm$ 0.47    &  0.52  $\pm$     0.11   & 198.18   $\pm$   79.27    &  10.98  $\pm$    0.08      &    11.22  $\pm$  0.40  \\

\hline
\noalign{\smallskip}
         
ID    &   RA      &    DEC    &       Redshift            & $(I-J)$ &  $J$    &  $\log \langle I_{\rm e} \rangle /(L_{\odot}pc^{-2})$         &   \re                    &      $n$               &       $q$               & $\sigma_{\rm e}$           &$\log M_\star$/\msun      & $\log M_{\rm dyn}$/\msun \\ 
      &                &               &                        &         &  mag         &      &         kpc            &                        &                      &  \kms                   &                         &                                   \\
\hline

12177  & 53.052200 &-27.774770 &1.6078   &  2.42   & 21.63   &  3.68  $\pm$       0.11  &   2.01  $\pm$    0.23         &    1.62  $\pm$    0.20      &   0.54  $\pm$    0.02        &     318.46  $\pm$  131.15      &  11.13 $\pm$     0.12         &    11.55 $\pm$     0.36    \\ 
11827  & 53.044943 &-27.774395 &1.6102   &  2.52   & 20.93   &  2.90  $\pm$       0.26  &   6.76  $\pm$    2.01         &    2.33  $\pm$    0.42      &   0.59  $\pm$    0.05        &     227.67  $\pm$   87.09      &  11.49 $\pm$     0.07         &    11.76 $\pm$     0.36    \\   
21853  & 53.062822 &-27.726461 &1.6110   &  2.12   & 21.37   &  3.97  $\pm$       0.07  &   1.69  $\pm$    0.11         &    3.38  $\pm$    0.22      &   0.95  $\pm$    0.02        &     212.09  $\pm$   64.22      &  11.13 $\pm$     0.13         &    11.05 $\pm$     0.26    \\   
25972  & 53.104571 &-27.705422 &1.6136   &  2.47   & 21.81   &  3.65  $\pm$       0.11  &   1.93  $\pm$    0.22         &    2.88  $\pm$    0.35      &   0.90  $\pm$    0.03        &     209.87  $\pm$   86.64      &  11.07 $\pm$     0.10         &    11.12 $\pm$     0.36    \\

\hline

\label{tab:summary_KCS}

\end{tabular}

\begin{scriptsize}
\begin{minipage}{18cm}

  {\sc Notes.} --- {  Galaxy IDs for \xmm\ and \cl\ come from our
    $H_{\rm F160w}$ catalogs (e.g., Chan et al sub.), whereas for
    \xcs\ from the $z_{\rm F850lp}$ catalog. Galaxies in clusters are
    listed in increasing redshift order matching that of
    Figure~\ref{fig:spectra}.} For \xmm\ the $H$-band corresponds to
$H_{\rm F160w}$, whereas the $(z-H)$ colors to
$(z_{\rm F850lp}-H_{\rm F160w})$. For \xcs\ $J$ corresponds to
$J_{\rm MOIRCS}$, whereas the $(z-J)$ colors to
$(z_{\rm F850lp}-J_{\rm MOIRCS})$. For \cl\ $J$ band is
$J_{\rm F125W}$ and $(I-J)$ corresponds to
$(I_{\rm F814W}-J_{\rm F125W})$, see CMDs in
Figure~\ref{fig:cmd}. Magnitude and colors are extinction
corrected. The surface brightness within the effective radius
$\log \langle I_{\rm e} \rangle$ is in rest-frame $B$-band.
$R_{\rm e}$ is the circularized effective radius, $n$ the S\'ersic
index, $q$ the axis ratio, and $\sigma_{\rm e}$ the stellar velocity
dispersions within \re. Only galaxies for which we measured stellar
velocity dispersion, the "dispersion sample'', are listed here. We
remind the reader that with our cosmology, 1\farcsn\ corresponds to
8.43 kpc, 8.45 kpc, and 8.47 kpc, at the mean redshift of our
overdensities ($z=1.39$, $z=1.46$, and $z=1.61$), respectively.

\end{minipage}
\end{scriptsize}

\end{center}
\end{tiny}
\end{table*}
%*****************************************************************************

\subsection[]{KMOS observations}
\label{subsec:kmos_data}

Observations were prepared with the KMOS Arm Allocator (KARMA,
\citealt{Wegner2008}) allocating 19-20 arms to objects and 1$-$3 arms
to faint stars, which were used to monitor both the point-spread
function (PSF) and the photometric conditions during the observations;
when more than one star was allocated, we used arms corresponding to
different KMOS detectors.
The total number of galaxies in the {\it red sequence} of the CMDs
within the limits described in Section~\ref{subsubsec:target_sel} that
were selected for KMOS observations of \xmm, \xcs, and \cl\ is 56.

Observations of the cluster galaxies were obtained between 30 Oct-16
Nov 2013, 6-19 July 2014, 19-21 October 2014, 17-19 September 2015,
10-12 October 2015, and 7-10 September 2016 using the KMOS $YJ$ band
filter covering the wavelength interval $1-1.36 \rm \mu m$.
Data for \xcs\ and \cl\ were taken with a standard
object-sky-object nodding pattern in which each on-source frame has an
adjacent sky exposure.  For \xmm\ we used a technique developed
for very crowded regions, in which we alternate IFUs on sky and
objects allowing to collect 100\% of the time on-source, but in half of
the objects compared to the two other clusters.  With two of those
allocations we obtain the same number of objects we observed in the
other two clusters.
Our single exposure times range from 300 s for \xcs\ and \cl\ to 450 s
for \xmm.  Each exposure was dithered by 0.1 $-$0.6\farcsn\ to improve
bad pixel rejection from the final extracted spectra.  The median
integration time per target is $\sim18$ hours on-source for our first
mask in \xmm\ and $\sim 10$ hours on-source for the second mask,
$\sim19$ hours on-source for objects from \xcs, and $\sim 16$ hours
on-source for objects from \cl.
We further apply additional quality cuts (i.e., seeing FWHM
$<1$\farcsn), such that the final exposure varies on a galaxy by
galaxy basis. The actual exposure time for the first mask of
\xmm\ ranges between 14 and 21 hours on source, between 5 and 16 hours
on source in the second mask, between 18-20 hours for \xcs, and between
6 and 17 hours for \cl.

Telluric stars, of spectral type O or A0, were observed as standard
calibrations.

\subsection[]{KMOS data reduction}
\label{subsec:data_red}

Data reduction was performed using a combination of the Software
Package for Astronomical Reductions with KMOS pipeline tools ({\tt
  SPARK}; \citealt{Davies2013}) and custom Python scripts 
(\citealt{Mendel2015}, Mendel et al, in prep).
{\tt SPARK} tools were used for bad pixel mask creation, flat fielding
and wavelength calibration. Science frames were corrected for a
channel-dependent bias drift using reference pixels on the perimeter
of each detector prior to reconstructing the data cubes. The
illumination correction was performed by using the observed sky-line
fluxes on a frame-by-frame basis. 
Sky subtraction was performed in two steps. We first performed a simple
A-B subtraction, then applied a second-order correction for the
residuals by stacking the spectra per detector. This second-order
correction effectively accounts for both the variability of OH lines
flux and system flexure.

We performed a 1D optimal extraction \citep[e.g.,][]{Horne1986} for the
targeted galaxies and used the \hst\ images in a band close to our
KMOS observations to describe how the galaxy flux is distributed
within the IFU (see \citealt{Mendel2015}, Mendel et al, in prep).

In summary, the center of the galaxy in each KMOS frame within an
acquisition was aligned to the \hst\ postage stamp position. The
shifts were derived from the dither pattern applied in the
observations and some additional centering accounting for the KMOS
positioning inaccuracy ($\sim$0\farcs2).  \hst\ images were convolved
to match the PSF measured from reference stars in each exposure.
We used the segmentation maps derived for the source catalogs to mask
neighboring objects and help optimize the extraction.
The spectra were extracted by combining the galaxy flux in the KMOS
IFU within the half-light radius of the galaxy, following a weighting
scheme and rejection criteria.
As an intermediate step, a telluric correction using both telluric stars
and model atmosphere through the code {\tt MOLECFIT}
\citep{Kausch2014} was applied, as well as a flux calibration.

We constructed 100 bootstrap realizations of the final 1D spectrum by
randomly selecting (with replacement) from the input frames; those
were used to estimate the uncertainties on the extracted spectra.

\section[]{Galaxy properties}
\label{sec:gal_prop}

\subsection[]{Structural parameters}
\label{subsec:sizes}

Structural parameters are derived in \citet{Chan2016} and Chan et al. (in
prep).  We use an adapted version of {\tt GALAPAGOS}
\citep{Barden2012}, which includes the two-dimensional light profile
modeling from {\tt GALFIT v.3.0.5} \citep{Peng2010a}. We perform
two-dimensional \citet{Sersic1968} fits, accounting simultaneously for
neighboring objects and deriving the PSF from bright stars in the
field.
We construct catalogs of structural parameters including semi-major
axis effective radii, $a_{\rm e}$, S\'ersic index $n$, and magnitude
in different ACS and WFC3 bands (see \citealt{Chan2016} and Chan et
al. sub.).
For each overdensity we select parameters derived in the observed band
closer to the rest-frame $B$-band of the local Coma FP
\citep[e.g.,][]{Jorgensen2006}.
For \xcs\ there is no available band exactly corresponding to the
rest-frame $B$-band, therefore we resort to using the ACS
$z_{\rm F850lp}$ effective radii and apply a wavelength-dependent
correction to the bluer effective radii, following the prescriptions
described in Section~5.1 of \citet{Chan2016}.

For \xmm, simulations show that our method produces measurements of
the semi-major axis $a_{\rm e}$ with a systematic bias of -0.4\% and a
$1\sigma$ dispersion less than 31\% for objects with $Y_{\rm F105w}$
surface brightness brighter than $22.75$ \mass.
Similarly, for \xcs, we recover $a_{\rm e}$ with a systematic bias of 1\%
and a $1\sigma$ dispersion less than 49\% for objects with
$z_{\rm 850lp}$ surface brightness brighter than $23.25$ \mass.  For
\cl, we recover $a_{\rm e}$ with a systematic bias of 1\%
and a
$1\sigma$ dispersion less than 5\% for objects with $J_{\rm 125w}$
surface brightness brighter than $20.75$ \mass\ (see \citealt{Chan2016}
and Chan et al., sub.)

Effective radii are circularized following $R_{\rm e}= \sqrt{ab}$,
where $a$ and $b$ are the semi-major and minor axis, respectively. For
our sample median errors on \re\ range from 0.79, 1.01, 0.22 kpc,
corresponding to 31\%, 37\%, and 11\% for \xmm, \xcs, \cl,
respectively.

{  Appendix~A of Chan et al., sub. shows} that our structural parameters for the
galaxies in \cl\ are consistent with publicly available measurements
by \citet{vanderWel2014} in the GOODS-S deep field (median difference
of $-0.029$ dex, and $1\sigma$ dispersion of 0.046 in
$\log R_{\rm e}$).

We derive absolute magnitudes in the rest-frame $B$-band Johnson
following Equation~2 of \citet{Hogg2002}, which includes the
apparent magnitude of our galaxies in the observed band, the distance
modulus at the redshift of each galaxy, and the K-corrections from
observed to rest-frame band \citep[e.g.,][]{Oke1968,Hogg2002}.
The K-corrections account for the factor $(1+z)$ related to the flux change with
redshift.

Given the limited amount of available photometric bands for our
sample, we resorted to calculating K-corrections for \citet{Maraston2005}
simple stellar population models (SSPs) with solar metallicity and
spanning the color range of our red-sequence galaxies.
Following Equation~2 of \citet{Hogg2002} the K-correction from one
observed band to the rest-frame $B$-band can be obtained knowing the
absolute magnitudes in $B$ band of the SSPs, their apparent magnitude
in the observed band and the distance modulus, as a function of
redshift and SSP observed band.  We obtain the above-mentioned
quantities for \citet{Maraston2005} SSPs using the code {\tt EzGal} of
\citet{Mancone2012}.
We fit quadratic functions between the model's K-corrections and their
colors and derive the appropriate K-correction for our galaxies; the
scatter of the relation was added in quadrature to the uncertainties
of the magnitudes. Median values of the K-correction from $Y_{\rm
  F105w}$ to $B$ band for \xmm\ are -0.96 mag, from $z_{\rm 850lp}$ to
$B$ band  0.06 mag for \xcs, and from $J_{\rm F125w}$ to $B$ band 
-1.36 mag for \cl.
We
find similar K-corrections if we adopt \citet{Bruzual2003} SSPs with
solar metallicity.

We derive the mean surface brightness within \re, $\langle I_{\rm e}
\rangle$ by dividing the total luminosity in the rest-frame $B$-band
by $2\pi R_{\rm e}^2$ (with $R_{\rm e}$ in pc), assuming a $B$-band solar magnitude in AB of
$M_{\odot,B}=$5.36 from Table~1 of \citet{Blanton2007}. 

All the parameters for the galaxies that are part of our dispersion sample are
summarized in Table~\ref{tab:summary_KCS}.

\subsection[]{Stellar masses}
\label{subsec:stellarmasses}

We use the stellar masses \mstar\ from \citet{Chan2016} and Chan et al
sub.. These masses are derived using an empirical $M_\star/L$ -
color relation \citep[e.g.,][]{Bell2001,Bell2003} calibrated on the
multi-band photometric catalogs of the NEWFIRM Medium-Band Survey
\citep[NMBS,][]{Whitaker2011} based on \citet{Bruzual2003} templates,
an exponentially declining star-formation history and a Chabrier
\citep{Chabrier2003} initial mass function (IMF). 
We correct from aperture to total \mstar\ using the best-fit S\'ersic luminosity.
Typical uncertainties range from $\sim 0.05-0.13$ dex, and include
photometric uncertainties in color and luminosity, as well as the
scatter in the derived $M_\star/L$ - color relation, but do not
include systematics like different IMF. Stellar masses for the
dispersion sample are listed in Table~\ref{tab:summary_KCS}.
The stellar masses we use are consistent with catalogs available in
the literature (see \citealt{Chan2016} and Chan et al sub.).
For \xmm\ we find a median difference $\Delta \log M_\star =  \log
M_{\star,\rm literature}-\log
M_{\star,\rm our}=0.03$ dex
and $1\sigma$ scatter of 0.09 dex with respect to $\log M_\star$
measurements by \citet{Delaye2014}, and for \xcs\ $\Delta \log
M_\star= -0.08$ dex and $1\sigma$ scatter of 0.14 dex from the \mstar\
measure in the same paper.  For \cl\ we find
$\Delta \log M_\star= -0.05 $ and $1\sigma$ scatter of 0.08 dex
compared to \mstar\ derived by \citet{Momcheva2016} and $\Delta \log M_\star=-0.05 $ and $1\sigma$ scatter of 0.06 dex
compared to \mstar\ derived by \citet{Santini2015}.

%*****************************************************************************
%                FIGURE 3
%*****************************************************************************
\begin{figure*}[!]
\centering
\includegraphics[width=0.9\textwidth]{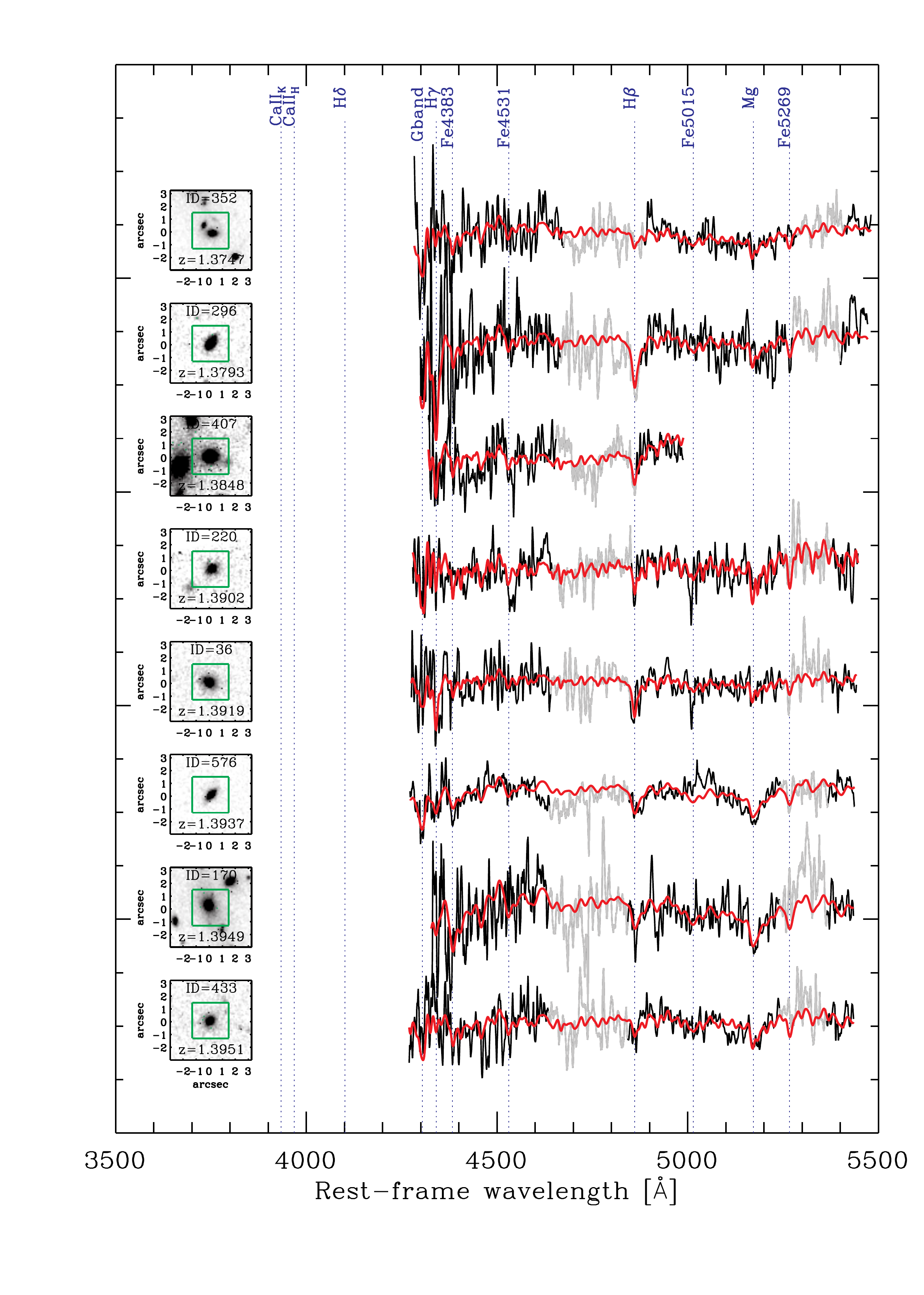}
\caption[] {In black we show {  continuum-normalized}, inverse-variance
  smoothed galaxy spectra with a window of {  7} pixels, ordered by
  redshift and shifted to rest-frame wavelengths. Kinematics fits are
  in red. In gray are the regions with strong telluric features or OH
  residuals that were excluded from the fit. The most prominent
  spectral features are labeled at the top of the figure and indicated
  by the dotted vertical lines. For each object we indicate IDs and
  {  redshifts} from Table~\ref{tab:summary_KCS} and show the \hst\
  postage stamps (either WFC3/$J_{\rm F125w}$ or ACS/$z_{\rm F850lp}$)
  of each galaxy of 6\farcsn\ side; the KMOS IFU is drawn in green.  }
\label{fig:spectra}
\end{figure*}

\addtocounter{figure}{-1}
\begin{figure*}[!]
\centering
\includegraphics[width=0.9\textwidth]{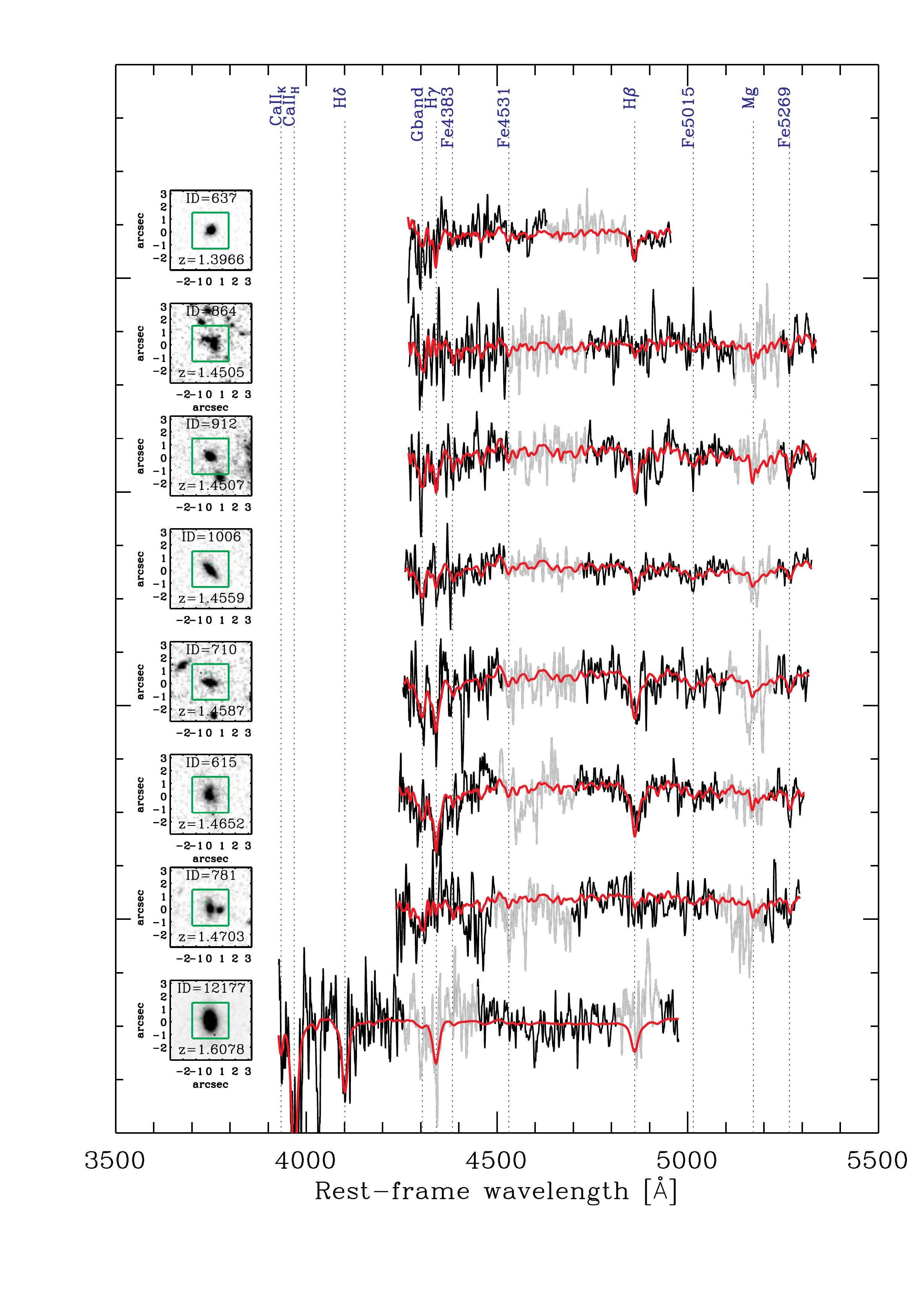}
\caption[$-$ Continued]{$-$ Continued}
\label{fig:spectra2}
\end{figure*}

\addtocounter{figure}{-1}
\begin{figure*}[!]
\centering
\includegraphics[width=0.9\textwidth]{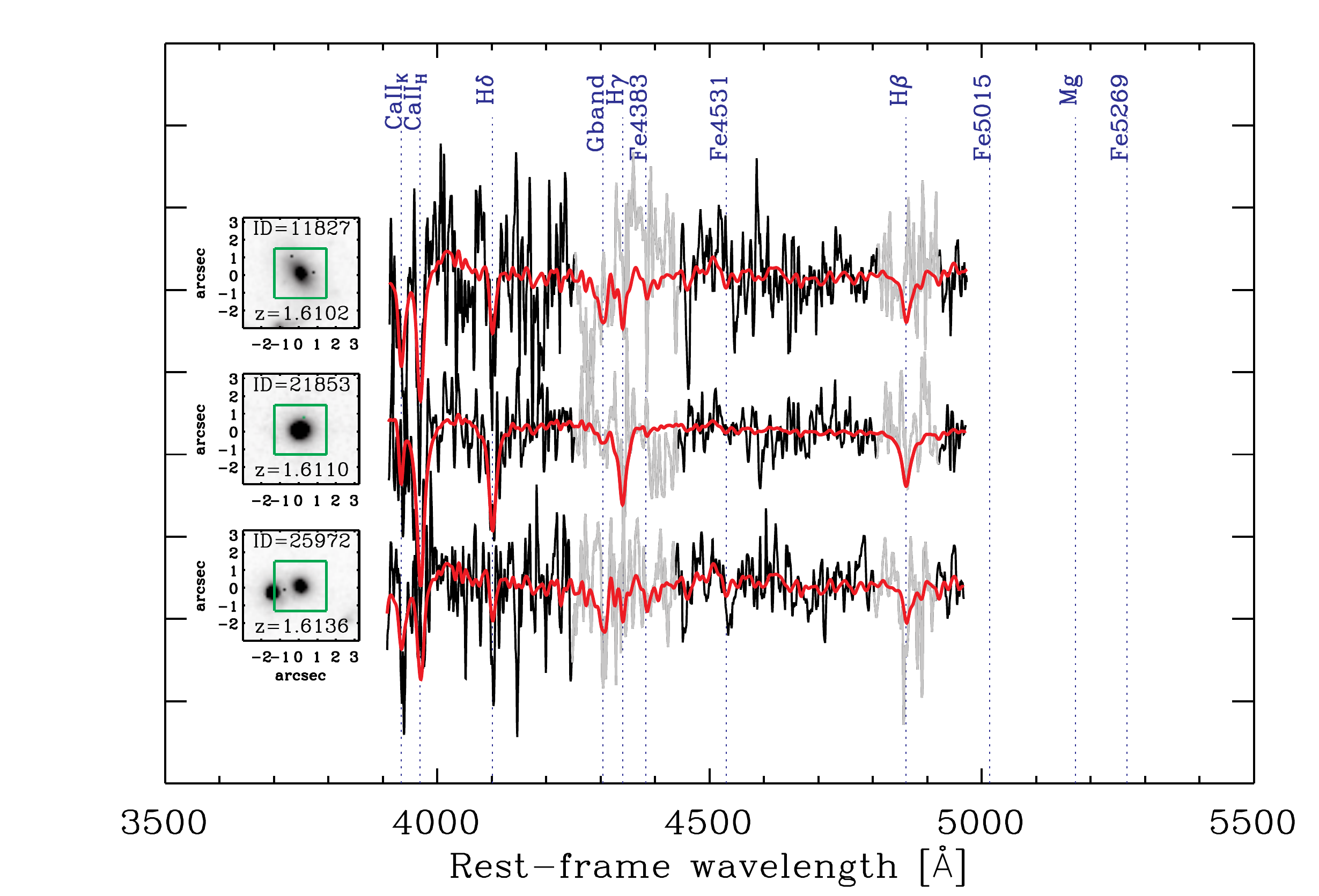}
\caption[$-$ Continued]{$-$ Continued}
\label{fig:spectra3}
\end{figure*}
%*****************************************************************************

\subsection[]{Stellar velocity dispersions}
\label{subsec:sigma}

We measure stellar velocity dispersions $\sigma$ with the Penalized
PiXel-Fitting method ({\tt pPXF}) of
\citet{Cappellari2004}, \citet{Cappellari2017}, by using as templates the
\citet{Maraston2011} stellar population models (SSPs) based on the
ELODIE v3.1 stellar library \citep{Prugniel2001, Prugniel2007}.
The ELODIE-based templates cover the wavelength range between
$3900-6800$ \AA, which matches the rest-frame wavelength of our KMOS
data, and have higher resolution (FWHM of 0.55 \AA) compared to the
median rest-frame KMOS FWHM ($\sim 1.5$\AA).
We did not use the more commonly used \citet{Maraston2011} SSPs based
on the MILES library (\citealt{SanchezBlazquez2006} with resolution
2.54 \AA\ as derived in \citealt{Beifiori2011b}) to allow us to derive
stellar velocity dispersion also in less massive objects. We tested
that our results are independent of the template library we used.

The KMOS spectra have a wavelength-dependent line spread function as
measured from the sky lines. Therefore, they are smoothed with a
variable kernel to match the maximum FWHM in each IFU before
fitting the kinematics. The use or not of this smoothing does not change our
measurements.
Consistently, \citet{Maraston2011} ELODIE-based templates are broadened to match
the new KMOS resolution.  Before the {\tt pPXF} fit, the KMOS spectra
are cross-correlated with a 1 Gyr old \citet{Maraston2011} template to
get an initial estimate for the redshift of the galaxy.

We use additive polynomials of low order, generally 2$-$3, to account
for uncertainties in the sky subtraction, {  which is one of the
  dominant factors of systematics in our data reduction, as well as
  any residual template mismatch. We test that the fit does not change by
  using polynomials of different orders (2$-$5): the large scale
  variations included by the additive polynomial have a minimum impact
  on the scales of our line widths. The use of multiplicative
  polynomials of low order, together or instead of additive
  polynomials give results consistent within the errors. During the
  fit, we also exclude regions with strong telluric and
OH line features.}

Stellar velocity dispersions are measured using the typical rest-frame
optical features, such as Ca H\&K, G-band, Balmer lines, and Mg
\citep{Bender1990,Bender1994}, depending on the cluster redshift; see
e.g. Figure~\ref{fig:spectra} for some examples of typical absorption
lines.
We measure $\sigma$ for 19 objects that constitute the dispersion
sample; our fits are shown in Figure~\ref{fig:spectra}, and the
derived parameters are summarized in Table~\ref{tab:summary_KCS}.
From our original sample, we mostly retained objects with measured
stellar velocity dispersion at the cluster redshift, with $S/N>5$,
{  values of \sigmae$<500$ \kms,} and where
the absorption line features were clearly visible after a visual
inspection of the fit.

We estimate uncertainties on $\sigma$ using the 100 bootstrap
realizations of the extracted spectra and recomputing the stellar
velocity dispersion for each one of those.  The errors on \sigmae\ are
the standard deviation of the distribution of the measurements of the
bootstrap realizations. Typical errors are of the order of
$\sim 11-40$\% for galaxies with a typical $S/N \sim 5-12$ per
angstrom.

%*****************************************************************************
%                Figure 4
%*****************************************************************************
\begin{figure*}
\centering
\includegraphics[width=\textwidth]{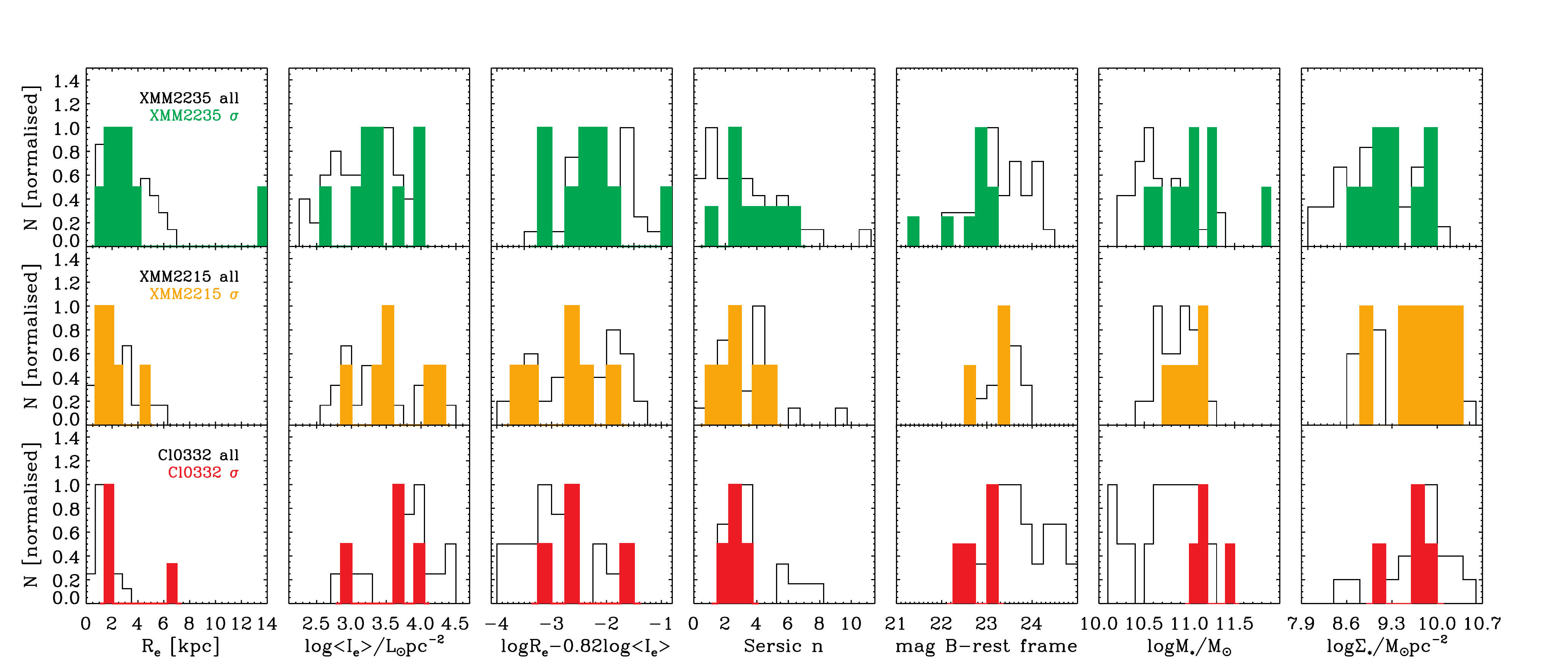}
\caption[] {Normalized distribution of galaxy properties for the full
  red-sequence sample (black) and for the sample for which we measured
  \sigmae\ (filled histograms).  From top to bottom we see \xmm, \xcs,
  \cl, in green, orange, red, respectively. From left to right:
  circularized effective radii, surface brightness, combination of
  size and surface brightness as in the y-axis of the FP, S\'ersic
  index, $B$-band magnitude, stellar mass, and surface mass density. 
  The sample with \sigmae\ spans a similar range of galaxy
  properties as the underlying population on the red sequence, though
  the limited statistics makes the sample  incomplete mostly at faint
  magnitudes and low stellar masses. 
}
\label{fig:galprop}
\end{figure*}
%*****************************************************************************

We perform tests to explore the robustness of our measurements by
analyzing different wavelength ranges of the spectrum. {  This
  provides an assessment of the systematic effects arising from the
  derivation of stellar velocity dispersions from different absorption
  lines at different redshift. We repeated the fitting by considering
  only the "blue" and the "red" regions of our spectra and their
  bootstrap realizations (see Appendix~\ref{sec:sigma_test} for
  details). For the two overdensities \xmm\ and \xcs\ where this test
  could be performed, the systematic offsets
  $\Delta \log \sigma_{\rm e}$ between the full spectrum fit and the
  two sub-regions are usually $<10$\%, and smaller than typical
  uncertainties on \sigmae.}

The KMOS 1D spectra are extracted within one $R_{\rm e}$. Because we
are background dominated the optimal extraction effectively makes our
stellar velocity dispersions luminosity weighted within \re, similarly
to the measurements provided in the local universe. For this reason no
aperture correction is needed, and we will therefore use the notation
$\sigma_{\rm e}$ for $\sigma$ throughout the paper.

\subsubsection[]{Dynamical masses}
\label{subsubsec:dynamicalmasses}

Dynamical masses \mvir\ are derived following the prescriptions of
\citet{Cappellari2006} by combining size and stellar velocity
dispersion measurements as $M_{\rm dyn}=\beta(n) R_{\rm e}\sigma_{\rm
  e}^2/G$, and assuming a S\'ersic index dependent virial factor
$\beta(n)$ (see \citealt{Beifiori2014} for a similar approach).
Typical uncertainties range from $\sim0.10-0.40$ dex.  The effect of
 change of the dark matter fraction
cannot be accounted in this simple mass estimator.
Moreover, another source for uncertainties is the unresolved rotation,
which is not accounted for by our method.  In fact, several works have
shown that at both intermediate and high redshift, rotational support
has a increasingly large contribution (\citealt{vanderMarel2007} at
$z\sim0.5$, and \citealt{Newman2015}; \citealt{Hill2016};
\citealt{Belli2017}; Mendel et al. in prep at $z>1.4$). 
Our sample, in particular, \xmm\ and \xcs, includes mostly objects
with S\'ersic index $>2.5$ where the impact of rotation is expected to
be small based on the findings of \citet{Belli2017}. Their Figure~7
shows that there is no trend in the ratio between stellar-to-dynamical
mass as a function of the axis ratio, whereas for disk galaxies with
S\'ersic index $<2.5$ there is a clear trend matching the expectations
for an increasing $V/\sigma$ at high redshift. For \cl\ half of the
objects could potentially be affected having in general lower S\'ersic
index compared to \xmm\ and \xcs\ (see Figure~\ref{fig:galprop}).

\subsection{Distribution of galaxy properties and selection effects}
\label{subsec:gal_properties}

Our sample does not appear to be systematically biased towards bluer
galaxies as found in previous kinematics samples at $z>1.4$
\citep[e.g.,][]{vandeSande2014} for \xmm\ and \cl; however some bias
could be present for \xcs\ (see Figure~\ref{fig:cmd}).

The unfilled histograms in Figure~\ref{fig:galprop} show the
distribution of circularized effective radii, effective surface
brightness, a combination of size and surface brightness, S\'ersic
indices, magnitudes, stellar masses, and surface mass density of the
red-sequence objects of the three clusters, whereas the filled
histograms show the distribution of our dispersion sample.  Surface
brightness and radii have similar distributions as the full sample in
particular for \xmm. However, we mostly have successful
$\sigma_{\rm e}$ measurements in the brightest and more massive
galaxies.

A Kolmogorov-Smirnov test shows that the probability that the
dispersion sample and the full red-sequence sample are drawn from the
same parent distributions in magnitude and stellar mass is $\sim 1$
and $13$\% for \xmm, $\sim 24$ and $55$\% for \xcs\, and $1$ and $3$\%
for \cl, respectively.  We are mostly unable to measure the stellar
velocity dispersion in faint objects.

We also perform a k-sample Anderson-Darling test \citep{Scholz1987} on
the same parameters of Figure~\ref{fig:galprop} to assess whether the
tail of the cumulative distributions can affect our results. We find
that the null hypothesis that the dispersion sample and the full
red-sequence samples are drawn from the same distribution in magnitude
and stellar mass can be rejected at 1\% and 5\%, respectively for
\xmm, $\sim 20$\% for \xcs, and 2\% and 5\% for \cl. For other
parameters such as \re, $\log \langle I_{\rm e} \rangle
/(L_{\odot}pc^{-2})$, and S\'ersic index the null hypothesis cannot be
rejected for \xmm, and \xcs\footnote{We note that for \xcs\ the null
  hypothesis cannot be rejected also considering the distribution of
  \mstar.}. For \cl\ the distribution of \re\ is not compatible with
the null hypothesis with a p-value of 20\%.

We therefore conclude that our observations are not representative of
the whole red-sequence sample in terms of stellar mass, or magnitude
for the three overdensities in our sample: for this reason in
Section~\ref{subsec:FP_m_l_z} we make a cut to a common stellar mass
limit.

{  We assess the selection effects and the success rate for the
  three clusters by deriving selection probabilities $\rm P_{\rm s}$
  for each galaxy with a method similar to that used by
  \citet{Saglia2010} (see Appendix~\ref{sec:FP_m_l_only_selection} for
  details).  In summary, $\rm P_{\rm s}$ are calculated accounting for
  the completeness in the measured stellar velocity dispersion,
  rescaled by the ratio of spectroscopically confirmed members over
  the targets we observed with KMOS. The inverse of $\rm P_{\rm s}$ are used as
  weights in the fits performed in Section~\ref{subsec:FP_m_l_z} to
  assess whether the use of our dispersion sample could bias our
  results.}

\section[]{Local and intermediate redshift comparison samples}
\label{sec:local}

\subsection[]{{  The Coma cluster}}
\label{subsec:coma}

As a local comparison sample we use the Coma cluster with a mass
$M_{200}=(1.6\pm0.4)\times 10^{15}$\msunh\ (rescaled mass from
\citealt{Lokas2003}, see also Figure~\ref{fig:cluster} and
Section~\ref{subsec:KCS_sample_paper}), which is usually adopted as a
reference cluster for FP and scaling relations studies (see e.g.,
\citealt{Jorgensen2006}, \citealt{Thomas2007}, \citealt{Houghton2013},
\citealt{Cappellari2013c}).

The photometry used to derive the local FP relation by
\citet{Jorgensen2006} is not public, therefore we adopt the structural
parameters and magnitudes from Table~A2 of \citet{Holden2010}. The
photometric parameters come from the surface brightness analysis to
the SDSS $g$-band images by \citet{Holden2007}, who transformed them
from SDSS bands to the rest-frame $B$-band.
The photometry of \citet{Holden2007} is consistent with that used in
the initial work on the Coma FP by \citet{Jorgensen1996} (see
description in Section~2.2.1 of \citealt{Holden2007} and Appendix~A2
of \citealt{Holden2010}) for the objects in common, but expanded to a
larger number of objects with stellar velocity dispersion from
\citet{Jorgensen1995} and \citet{Jorgensen1999} (which in total are
116 galaxies), making up a sample of 80 galaxies with both stellar
velocity dispersion and structural parameters.
In \citet{Holden2010} effective radii are given in the $g$ band; we
assume those to be comparable to rest-frame $B$-band radii throughout
the paper.

Dynamical masses are computed from the \sigmae, \re\ and S\'ersic {\it
  n} given in Table~A2 of \citet{Holden2010}, using the approach of
Section~\ref{subsubsec:dynamicalmasses}, as for our KCS sample. The
Coma cluster was targeted by the SDSS photometry, therefore we use
stellar masses from the SDSS catalog of \citet{Maraston2013}.
\citet{Maraston2013} stellar masses were derived from the SED fitting
of the five SDSS bands and were based on \citet{Maraston2009} stellar
population models and a \citet{Kroupa2001} IMF. We homogenize \mstar\
to a common Chabrier IMF and account for the difference in the stellar
population models used adopting the offsets provided in Table~B3 of
\citet{Pforr2012}\footnote{We note that the offset provided in
  Table~B3 of \citet{Pforr2012} are for galaxies at $z=0.5$. At the
  redshift of Coma this represents an upper limit on the offset we
  need to apply.}, which were derived for mock passive galaxies for
different population models \citep[see detailed description of the
method in][]{Beifiori2014}.
We test that similar results are obtained if we use other \mstar\
catalogs of SDSS galaxies \citep[e.g.,][]{Mendel2014}.
The rescaling of the stellar masses to the best-fit S\'ersic
luminosity has a minimal effect on the mass estimate we use.
The Coma sample is our local reference throughout the paper,
therefore we do not apply the mass cut we have in our KMOS sample at
$\log M_{\star}/M_{\odot}>10.5$ to minimize bias on the scaling
relations we derive.

We compare our Coma catalog with that of \citet{Cappellari2013c},
which includes effective radii in $K$-band and dynamical masses for a
$K$-band selected sample of 161 galaxies within a magnitude limit of
$M_{K}<-21.5$. This sample includes both early-type galaxies and
spirals ($\sim 10$\%). There are 90 objects in common between the two
samples, 66 of which have measured sizes from \citet{Holden2010}. We
find consistent distributions of sizes and masses between the two
samples.  For the objects in common, we find that sizes are mostly
consistent
($\Delta \log R_{\rm e} =\log R_{\rm e, Holden}-\log R_{\rm
  e,Cappellari}\sim-0.03$ dex and scatter 0.16 dex), whereas stellar
and dynamical masses are offset in respect to the masses of
\citet{Cappellari2013c}
($\Delta \log M=\log M_{\rm Cappellari} -\log M_{\rm our}$) of
$\sim$0.14 dex and $\sim-0.16$dex, respectively, where we included a
factor $\sim 0.08$dex in our measured stellar masses due to the
different aperture used deriving stellar and dynamical masses (see
Appendix~A of \citealt{Beifiori2014}).  Those differences will not
affect our conclusion.

The full Coma sample includes galaxies with a range of properties:
some of them may not descend from our high-redshift galaxies
\citep[e.g., ``progenitor
bias''][]{vanDokkum2001a,Valentinuzzi2010,Saglia2010,Poggianti2013,Carollo2013,Beifiori2014}.
Several studies have found a relation between the size and the age of
passive galaxies at intermediate and high redshift (e.g., \citealt{Valentinuzzi2010},
\citealt{Poggianti2013}, \citealt{Shetty2015}, \citealt{Belli2015}, but see also
\citealt{Fagioli2016} for different opinions), suggesting that the
selection of old galaxies could minimize the
descendant/progenitor mismatch. 
Following this idea, we select Coma galaxies whose ages
are older than the difference between the look-back times of Coma and
our highest redshift KCS overdensity (e.g., $> 9$ Gyr), following a
procedure similar to that of \citet{Beifiori2014}, \citet{Chan2016}.
Ages for Coma galaxies are derived from the line indices measurements
of \citet{Jorgensen1999} (their Table~4 contains measurements for 70
galaxies) using \citet{Maraston2005} stellar population models, and
following the method described by \citet{Saglia2010b}. The resulting
sample with an age $> 9$ Gyr consists of six galaxies.

%************************************************************************
% FIGURE 5
%************************************************************************
\begin{figure*}
\centering
\includegraphics[width=0.48\textwidth]{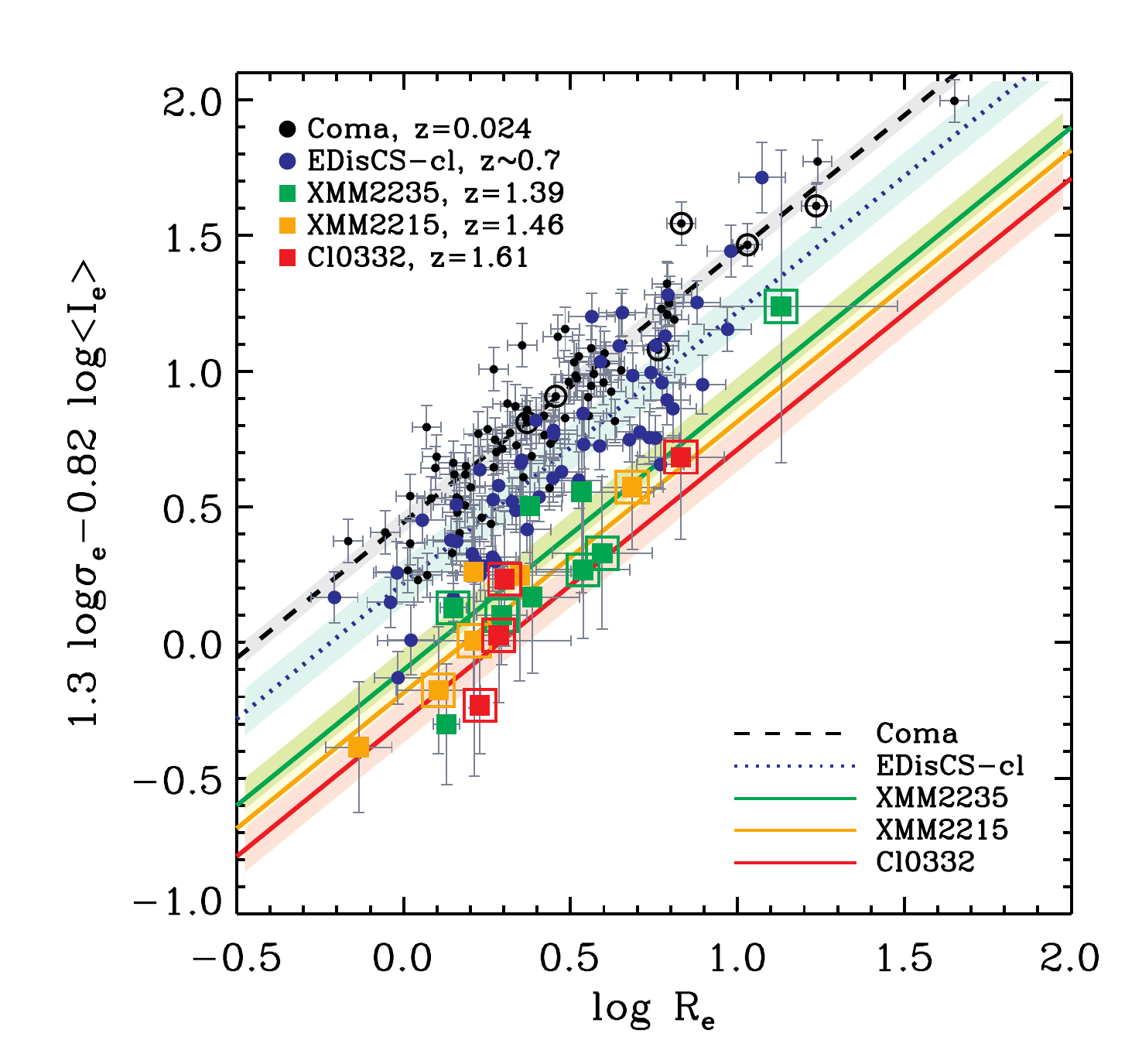}
\includegraphics[width=0.48\textwidth]{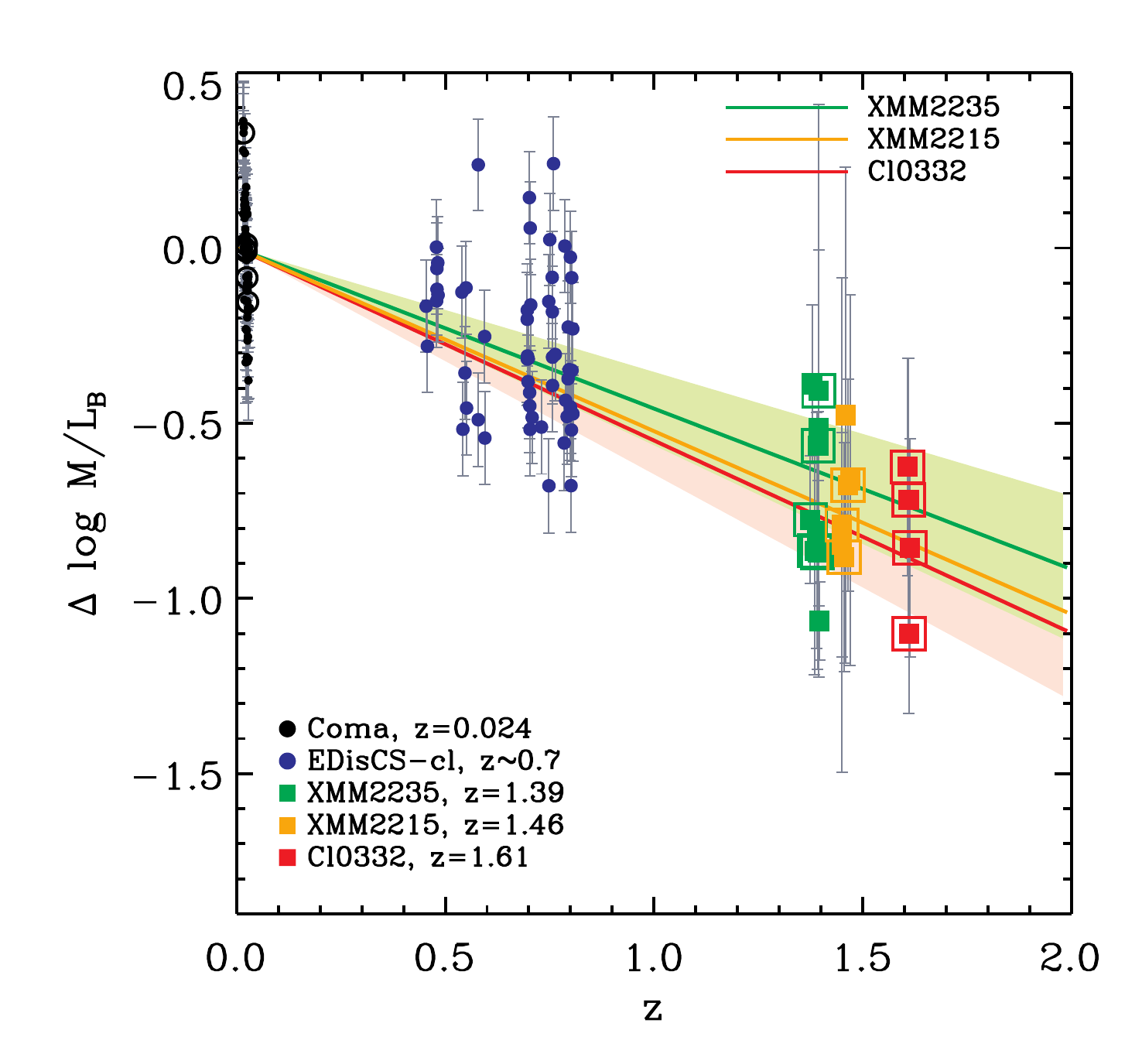}
\caption[] {Left panel: Edge-on projection of the FP. The Coma FP of
  \citet{Jorgensen2006} (black dashed line and black filled circles)
  and the FP from EDisCS clusters at $z\sim0.7$ of \citet{Saglia2010}
  (blue dotted line and blue filled circles) are included as
  reference. Open black circles show only Coma galaxies with an age
  $>$ 9 Gyr. KCS galaxies (green, orange and red filled squares)
  follow the FP scaling relation, but are offset (green, orange, and
  red solid lines) with respect to Coma and EDisCS.  Open squares for
  our KCS sample show galaxies for which $ \log M_\star/M_\odot>11$. Shaded
  regions show the $1\sigma$ scatter.  Right panel: Redshift evolution
  of $\Delta \log M/L_{B}$.  Each solid line (green, orange and red)
  shows the best-fit linear relation for Coma and the galaxies in each
  of the KCS overdensities with $ \log M_\star/M_\odot>11$ (excluding EDisCS
  from the fit, see text). Shaded regions show the $1\sigma$ errors on
  the slope. The shaded region for \xcs\ is covered behind that of
  \xmm, and \cl.}
\label{fig:fp_ml}
\end{figure*}
%************************************************************************

\subsection[]{{  The EDisCS-cluster sample}}
\label{subsec:ediscs}

The intermediate redshift comparison sample comes from the cluster
galaxies at median redshift $z\sim 0.7$ used for the FP work of the
EDisCS survey \citep[][see also Figure~\ref{fig:cluster} and
Section~\ref{subsec:KCS_sample_paper}]{Saglia2010}. The full
EDisCS-cluster sample includes 26 clusters or groups with redshift
between $0.4<z<0.9$ and velocity dispersion between 166 to 1080 \kms
\citep[][]{Milvang_Jensen2008,Saglia2010}. In this paper we 
consider a subsample of 10 clusters out of the 14 with  available \hst\
photometry, for which both S\'ersic fits and stellar masses
were derived (see Figure~\ref{fig:cluster}).  Catalogs include
structural parameters from \citet{Simard2009}, stellar velocity
dispersions from \citet{Saglia2010} and updated \mstar\ from Rudnick
et al. (sub) derived using the {\tt iSEDfit} software
\citep{Moustakas2013}, \citet{Bruzual2003} models and a Charbier IMF;
stellar masses were also rescaled for the missing flux using the
best-fit S\'ersic luminosity similarly to our KCS sample. Dynamical
masses are derived as in Section~\ref{subsubsec:dynamicalmasses}.
We select galaxies with $\log M_{\star}/M_{\odot}>10.5$, to be
consistent with the stellar mass limit of our KMOS sample. {  For
  some of the tests in Section~\ref{subsec:FP_m_l_z} we applied an
  additional mass cut $\log M_{\star}/M_{\odot}>11$ to match the mass
  limit of our highest redshift overdensity.}
Also in this case, we select a subsample of galaxies with an age $>3$
Gyr, to minimize progenitor bias. Ages for EDisCS galaxies were
collected from \citet{SanchezBlazquez2009}.
The resulting sample consists of 56 galaxies, which we will identify
as the EDisCS-cluster sample (``EDisCS-cl''), throughout the paper.

\section[]{Results}
\label{sec:FP_clusters}

\subsection[]{The Fundamental Plane of KCS galaxies}
\label{subsec:FP_zero_point}

We consider the edge-on projection of the fundamental plane as:

\begin{equation}
\log R_{\rm e}= a \log \sigma_{\rm e} +b \log \langle I_{\rm e} \rangle +c_z
\label{eq:fp}
\end{equation}

\noindent
where \re\ is in kpc, \sigmae\ in \kms,
$\log \langle I_{\rm e} \rangle $ is in \lsunpcsq, $c_z$ is the
redshift-dependent zero-point of the relation, and $a=1.30\pm0.08$ and
$b=-0.82\pm0.03$ come from the local $B$-band FP of
\citet{Jorgensen2006}.

Due to the limited number of objects in our sample, we assume the
local slopes are still valid at high redshift. {  We note that
  some variations of the coefficients were reported by previous work at
  $z\sim1$ (e.g., \citealt{Treu2005}, \citealt{vanderWel2005},
  \citealt{Renzini2006}, \citealt{Saglia2010},
  \citealt{Jorgensen2013})}.

We determine the zero-point of our KCS galaxies performing a
least-square fit using the {\tt MPFITEXY} routine
\citep{Williams2010}, by accounting for the errors in both
coordinates, no intrinsic scatter, and assuming the slopes fixed to
the local value. The {\tt MPFITEXY} routine depends on the {\tt MPFIT}
package \citep{Markwardt2009}.  Errors are obtained with a
Jackknife method. 

The left panel of Fig.~\ref{fig:fp_ml} shows the edge-on projection of
the FP for our KCS sample compared to the EDisCS sample
\citep{Saglia2010} and the Coma relation of \citet{Jorgensen2006}.
The zero-point evolves with redshift from the value 0.443 of Coma in
$B$ band \citep[][at $z=0.023$]{Jorgensen2006} to $0.22\pm0.02$ at the
median $z\sim0.7$ of EDisCS-cl to $-0.10\pm0.09$, $-0.19\pm0.05$,
$-0.29\pm0.12$ for \xmm\ at $z=1.39$, \xcs\ at $z=1.46$, and \cl\ at
$z=1.61$, respectively. We note that the zero-points are derived
including the full sample of galaxies from the three KCS clusters. The rms
scatter of the relations ranges from 0.08 for Coma to $\sim 0.15\pm
0.05$ for EDisCS-cl, to $\sim 0.09\pm 0.02$ for \xmm, $\sim 0.09\pm
0.02$ for \xcs, and $\sim 0.16\pm 0.07$ for \cl.

We test that the photometry we use for the Coma galaxies allows us to
recover the zero-point of Coma by \citet{Jorgensen2006} within the
errors of about $\sim 0.02$. We also test whether the use of Coma
galaxies with older ages affect our results.  We fix the coefficients
$a$ and $b$ of the fundamental plane of \citet{Jorgensen2006} and
derive the zero-point $c_z$ using subsamples selected with
increasingly larger age limit, up to our progenitor bias limit of $>9$
Gyr. Our test shows that the zero-point varies at most by 0.02,
suggesting that the use of the full Coma sample does not bias our
results; for this reason we can assume the same zero-point derived by
\citet{Jorgensen2006} throughout the paper.

%************************************************************************
%      TABLE 2
%************************************************************************
\begin{table*}
\begin{scriptsize}
\begin{center}
\caption{Evolution of $\Delta \log M/L_{B}$ with redshift.}
\begin{tabular}{c c c c c }
\hline
\hline
\noalign{\smallskip}
Case & Relation                                                                                             &  \xmm                   &    \xcs        &   \cl          \\
\noalign{\smallskip}
\hline
\noalign{\smallskip}              
 1      &$\Delta \log M/L_{B}= \eta\log(1+z)$, KCS, $\log M_{\star}/M_{\odot}>11$                                  & $-1.68\pm0.37$          &  $-1.91\pm0.25$    & $-2.10\pm0.37$           \\
{  2} &$\Delta \log M/L_{B}= \eta\log(1+z)$, EDisCS-cl+KCS, $\log M_{\star}/M_{\odot}>11$                        & $-1.33\pm0.13$          &  $-1.28\pm0.15$    & $-1.32\pm0.15$           \\
{  3} &$\Delta \log M/L_{B}= \eta\log(1+z)$, EDisCS-cl+EDisCS-cl, $\log   M_{\star}/M_{\odot}>11$ \&  $M_{200}$   & $-1.44\pm0.15$          &  $-1.18\pm0.16$    & $-1.32\pm0.21$           \\
 4      &$\Delta \log M/L_{B}= \eta\log(1+z)$, all  KCS                                                        & $-1.98\pm0.46$          & $-1.92\pm0.15$     & $-2.10\pm0.37$          \\
 5      &$\Delta \log M/L_{B}= \eta\log(1+z)$, KCS, $\log M_{\star}/M_{\odot}>11$ \& $\rm P_{\rm s}$                & $-1.65\pm0.41$          & $-1.93\pm0.23$     & $-2.17\pm0.41  ^{*}$    \\

\noalign{\smallskip}
\hline
\noalign{\smallskip}
\label{tab:cluster_fit}
\end{tabular}
\begin{scriptsize}
\begin{minipage}{17cm}
  {\sc Notes.} --- Evolution of $\Delta \log M/L_{B}$ for our { 
    five} test cases: 1) KCS galaxies with $\log
  M_{\star}/M_{\odot}>11$, {  2) $\log M_{\star}/M_{\odot}>11$
    galaxies part of the full sample of EDisCS-cl and KCS, 3) $\log
    M_{\star}/M_{\odot}>11$ galaxies part of the EDisCS-cl and KCS,
    but with EDisCS-cl sorted based on their $M_{200}$ to match that
    of the three KCS overdensities,} 4) the full sample of galaxies
  for which we derived stellar velocity dispersion in KCS, and 5) the
  KCS galaxies with $\log M_{\star}/M_{\odot}>11$ in which selection
  weights $\rm P_{\rm s}$ are applied. $^{*}-2.18\pm0.37$ is the slope
  we obtain if we apply the selection weights derived for the full
  GOODS-S structure.
\end{minipage}
\end{scriptsize}
\end{center}
\end{scriptsize}
\end{table*}
%************************************************************************

\subsection[]{The $M/L$ evolution with $z$}
\label{subsec:FP_m_l_z}

Under the assumption of homology (i.e,  that the
  coefficients $a$ and $b$ are constant with redshift, see
  Section~\ref{subsec:FP_zero_point}), the zero-point of the FP traces the
mean galaxy $\Delta \log M/L$. 
Therefore, we can convert the zero-point change into an evolution of
the $\Delta \log M/L$ ratio with redshift such that:

\begin{eqnarray}
\Delta \log (M/L)_{z}& = & \log (M/L)_{z}- \log(M/L)_{\rm Coma}
                         \nonumber \\ 
                  & = &(c_{z} -c_{\rm Coma})/b
\label{eq:m_l}
\end{eqnarray}

\noindent
where $c_z=\log R_{\rm e}- (a\log \sigma_{\rm e }+b \log \langle
I_{\rm e} \rangle)$ based on Eq.~\ref{eq:fp} and $c_{\rm Coma}$ is the
Coma zero-point of \citet{Jorgensen2006}. The error on the Coma
zero-point is not provided by \citet{Jorgensen2006}, therefore we
obtained it from our Coma sample with a Jackknife method as described
in Section~\ref{subsec:FP_zero_point}.

The right panel of Fig.~\ref{fig:fp_ml} shows the evolution of the
$\Delta \log M/L$ as a function of redshift.  The overdensities in our
sample have different properties and masses (see
Section~\ref{subsec:KCS_sample_paper}), therefore we could expect
different mean ages for our galaxies. For this reason, we separately
fit our KCS galaxies, using Coma as a reference value. The use of the
Coma cluster to define the zero-point is arbitrary being a
normalization term, and will not affect our analysis.  {  Our
  intermediate-redshift EDisCS-cl sample has significantly
  smaller error bars compared to our KCS clusters. In
  Table~\ref{tab:cluster_fit} we show that this drives the fit towards
  the slopes preferred by the EDisCS-cl sample. For this reason we
  resort giving the slopes derived fitting only Coma and KCS galaxies. }

  To make a fair comparison between different overdensities we
  consider the $\Delta \log M/L$ evolution for a sample with the
  same mass distribution in the three overdensities.  We apply a mass
  cut of $\log M_{\star}/M_{\odot}>11$ to the sample of \xmm\ and
  \xcs\ to match the minimum \mstar\ for the objects in \cl.  This
  mass limit is also known to provide an unbiased measure of the $M/L$
  evolution as discussed in \citet{vanderMarel2007} and
  \citet{vanDokkum2007}.
  We obtain $\Delta \log M/L_{B}=(-0.46\pm0.10)z$, $\Delta \log
  M/L_{B}=(-0.52\pm0.07)z$, and $\Delta \log M/L_{B}=(-0.55\pm0.10)z$,
  for \xmm, \xcs, and \cl, respectively. Error bars were estimated
  with a  Jackknife method.
  In \xmm\ and \xcs\ we were able to derive stellar velocity
  dispersions also in galaxies with $10.5<\log
  M_{\star}/M_{\odot}<11$.  If we include those in our analysis we
  obtain slightly steeper $\Delta \log M/L_{B}=(-0.54\pm0.13)z$,
  thought consistent within the errors, for \xmm\, and consistent
  $\Delta \log M/L_{B}=(-0.52\pm0.04)z$ for \xcs.

The $\Delta \log M/L$ evolution for the massive galaxies of \xcs\ and \cl\ is
consistent with the results by \citet{Saglia2010} and
\citet{vanDokkum2007}, but slightly steeper than
\citet{vanDokkum2003}, \citet{Wuyts2004}, \citet{Holden2005}.
The $\Delta \log M/L$ evolution for the massive galaxies of \xmm\ is smaller.
\citet{vandeSande2014} found similar small $\Delta \log M/L$ evolution, studying a
sample of field galaxies at $z\sim1.5$ and applying a correction for
the fact that galaxies in their sample are bluer than a representative
parent sample of quiescent galaxies from the 3DHST survey
\citep{Brammer2012,Skelton2014}. As shown in Figure~\ref{fig:cmd} our
sample does not appear to be biased towards blue objects in \xmm\ and
\cl, for \xcs\ we note that the majority of the objects are in the
bluer part of the red sequence.

We now express the evolution of $\log M/L$ as
$\Delta \log M/L_{B}= \eta \log(1+z)$, where $\eta$ is the slope of
the logarithmic dependence. This parametrization allows us to compare
the amount of $\Delta \log M/L$ evolution compared to the structural
evolution with redshift in our sample (see
Section~\ref{subsec:FP_m_l_str}).

{  We derive the evolution considering different combinations of the
  available data, which
  are summarized in Table~\ref{tab:cluster_fit}. Case 1) shows the
  results for only KCS galaxies with $\log M_{\star}/M_{\odot}>11$;
  case 2) $\log M_{\star}/M_{\odot}>11$ galaxies part of the full
  sample of EDisCS-cl and KCS; case 3) $\log M_{\star}/M_{\odot}>11$
  galaxies part of the EDisCS-cl and KCS, but with EDisCS-cl sorted
  based on their $M_{200}$ (see Figure~\ref{fig:cluster}) to match
  that of the three KCS overdensities\footnote{We split EDisCS-cl in
    three samples: $M_{200}>10^{15} h^{-1} M_{\odot}$ for \xmm,
  $M_{200}>10^{14} h^{-1} M_{\odot}$ for
    \xcs, and $10^{13} h^{-1} M_{\odot} <M_{200}<10^{14} h^{-1} M_{\odot}$ for \cl.}; case 4) we include
  also KCS galaxies with $\log M_{\star}/M_{\odot}>10.5$ in \xmm, and
  \xcs; and case 5) we use only KCS galaxies with $\log
  M_{\star}/M_{\odot}>11$ for which we apply the selection weights
  $\rm P_{\rm s}$ we describe in Section~\ref{subsec:gal_properties}
  and derive in Appendix~\ref{sec:FP_m_l_only_selection}. Error bars
  are estimated with a Jackknife method.

  In case 1 we find that the slopes $\eta$ are consistent within the
  errors for the three KCS overdensities, with a weak suggestion that
  the slope of \xmm\ is slightly flatter than that of \cl. 

  In case 2 and 3 we generally find flatter slopes, which are
  consistent between the three KCS overdensities, and smaller error
  bars.  We note that the actual value of the slopes in those two
  cases is fully driven by the EDisCS-cl data that have significantly
  smaller error bars than our KCS clusters. To avoid being biased
  towards the intermediate-redshift sample slopes we refrain from
  using the EDisCS-cl sample in the following analysis and include the
  galaxies in our plots only for visual comparison.}

{  In case 4} the derived slope $\eta$ for \xmm\ is steeper, but
still consistent within the errors with case 1). This is the result of
the inclusion of the low mass, low \sigmae\ objects in the fit.  Our
findings support the results of \citet{Treu2005} and
\citet{Renzini2006}, where they found a change of the $\Delta \log
M/L_{B}$ with galaxy mass, with steeper $\Delta \log M/L_{B}$ in lower
mass objects as result of recent star formation.

In case 5) we assess the effect of the sample selection, and fit the
$\Delta \log M/L$ evolution by applying the selection weights $\rm
P_{\rm s}$ we derive in Appendix~\ref{sec:FP_m_l_only_selection} to
the same sample we used in case 1). This allows us to up-weight the
objects whose selection weights are smaller by scaling down their
errors on the fitting procedure. We find values of $\eta$ consistent
within the errors with the case with no weights, {  with a weak
  tendency, given our error bars, to have flatter slopes for \xmm\
  compared to \cl, which instead become steeper. For \cl, the use of
  selection weights $\rm P_{\rm s}$ calculated within the KMOS FoV and
  within the all GOODS-S structure give consistent results.}  We note
that this technique we can only account for selection effects for the
sample that we targeted at the cluster redshift.

\section[]{Discussion}
\label{sec:discussion}

\subsection[]{Formation ages from the $M/L$ ratio evolution}
\label{subsec:FP_m_l_only}

Passive evolution models are described by a formation redshift
corresponding to the epoch of the last major star formation episode,
{  which allow us to translate} the $\log M/L$ and luminosity
evolution into a formation age.

{  In order to derive formation ages from the $\log M/L$ ratio of
  our KCS galaxies, we use tabulated $M/L$ of the SSP models of
  \citet{Maraston2005} with a \citet{Salpeter1955} IMF and solar
  metallicity.  We interpolate the models with a cubic spline to obtain
  a equally spaced $\log M/L$ grid in $\log \rm age$.}  We use solar
metallicity \citet{Maraston2005} SSPs, which is supported by both
spectroscopic and photometry results of passive galaxies at
high-redshift \citep[e.g.,][]{Mendel2015,Chan2016}.  In
Appendix~\ref{subsec:ssp_Z} we show the effect of using different SSPs
and metallicity assumptions.

{  We test that our $\log M/L$ are consistent with the the
  $\log M/L$ we derive by fitting a linear relation between the
  logarithms of the age, metallicity and $M/L$ of \citet{Maraston2005}
  SSPs with an age $\geq$1 Gyr and metallicity $\log Z/Z_{\odot}$
  between $-0.3$ to $0.3$ (total metallicity relative to solar) as
  done in \citet{Jorgensen2005}, \citet{Jorgensen2013}, and
  \citet{Jorgensen2014}.  In the following, we adopt the $\log M/L$
  obtained via interpolation of the SSPs because the relation between
  $M/L$ and $\log \rm age$ becomes strongly non-linear at young ages,
  which could potentially make our linear relation more uncertain.}

\begin{figure}
\centering
\includegraphics[width=\columnwidth]{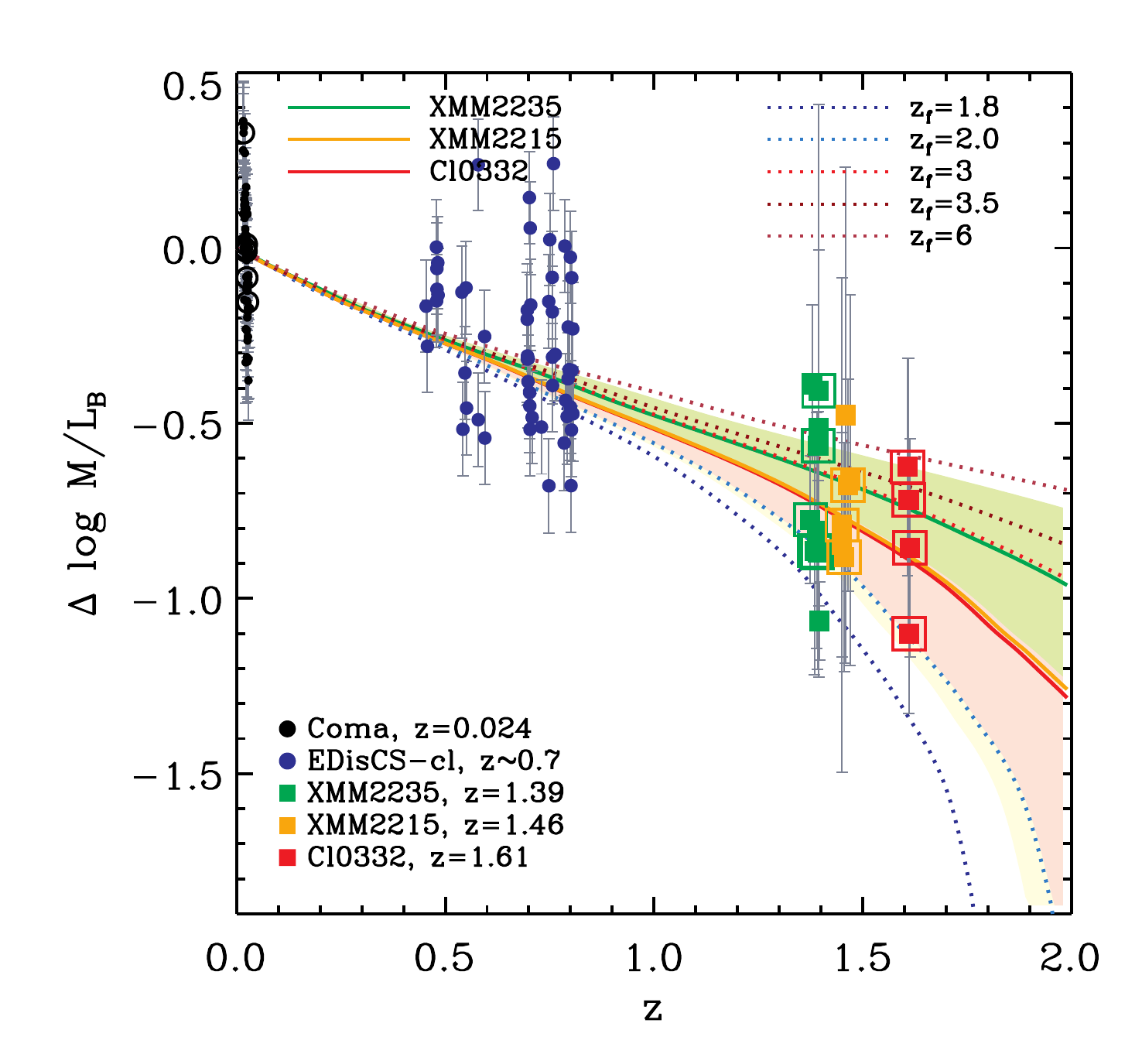}
\caption[] {Redshift evolution of the $\Delta \log M/L_{B}$. Symbols
  as in Figure~\ref{fig:fp_ml}.  Dotted lines correspond to the predictions by
  \citet{Maraston2005} SSPs for a Salpeter IMF, solar metallicity, and
  different formation redshifts.  Each solid line (green, orange and
  red) shows the SSP with formation age corresponding to that we
  derived for each cluster. Shaded regions show the $1\sigma$
  error on the slope. The shaded region for \xcs\ is covered behind that of
  \xmm, and \cl.
}
\label{fig:summary_FP1}
\end{figure}

Figure~\ref{fig:summary_FP1} shows \citet{Maraston2005}
SSPs for a range of formation redshifts $z=1.8-6$ (dotted lines).

In our analysis we use relative $M/L$; therefore changes in the
normalization of the IMF (i.e., changes in the low mass slope) will
not affect our conclusions.  We note that non standard IMF slopes { 
  (i.e, non$-$Salpeter)} would lead to different conclusion due to the
known degeneracy between IMF and formation redshift
\citep[e.g,][]{vanDokkum1998,Renzini2006,vanDokkum2007}. The influence
of the choice of the IMF will be discussed in a forthcoming paper.

{  We estimate the best-fit value and the 68\% confidence interval
  of the formation ages by integrating the posterior likelihood
  distribution of the only free parameter, i.e.  the formation age (or
  the formation redshift), given our data.  We assume a top-hat prior
  on the formation redshift $z=1.8-6$.  We sampled the likelihood on a
  grid of formation redshifts within the prior.}

The most massive galaxies in \xmm\ have a mean luminosity-weighted age
of $2.33^{+0.86}_{-0.51}$ Gyr, whereas the mean age of \xcs\ is $1.59^{+1.40}_{-0.62}$ 
Gyr, and $1.20^{+1.03}_{-0.47}$  Gyr for \cl. 
{  The ages of the three overdensities are consistent within the
  errors after accounting for the $\sim$0.5 Gyr difference in the age
  of the Universe between $z=1.61$ and $z=1.39$. There} is a hint,
albeit at low significance, that the ages of galaxies in \cl\ are
younger compared to ages of \xmm\ galaxies.
{  This weak suggestion is supported by our results from stacked
  spectra of the same galaxies, which will be described in Houghton et
  al., in prep.}

\cl\ is by far the lowest density environment in our survey and shows
ages of the order of field galaxies at similar redshift. If we compare
the ages measured from stacked spectra of massive galaxies at the
median redshift of $z=1.75$ in the KMOS VIRIAL survey
\citep[e.g.,][age of $1.03_{-0.08}^{+0.13}$ Gyr]{Mendel2015} we find
that the two measures agree within the errors, after accounting for
the 0.29 Gyr difference between the age of the Universe at $z=1.61$
and that at the median redshift of VIRIAL (see also
Section~\ref{subsec:cluster_vs_filed}). 

We have verified that our results are consistent within the errors
with ages obtained from the color evolution of SSPs compared to our
galaxies.

By including in the sample of \xmm\ objects with $10.5<\log
M_{\star}/M_{\odot}<11$ we derive slightly younger ages
$\sim1.63^{+0.39}_{-0.29}$ Gyr, suggesting a trend of age with \mstar,
as discussed in Section~\ref{subsec:FP_m_l_z}, such that lower mass
galaxies have younger ages compared to more massive objects. For \xcs\
we obtain consistent ages $\sim 1.62^{+1.30}_{-0.61}$ Gyr, which is
probably due to the number of bluer objects which entered our the
red-sequence sample.

%************************************************************************
% FIGURE 6
%************************************************************************
%************************************************************************

%************************************************************************
% FIGURE 7
%************************************************************************
\begin{figure*}
\centering
\includegraphics[width=0.8\textwidth]{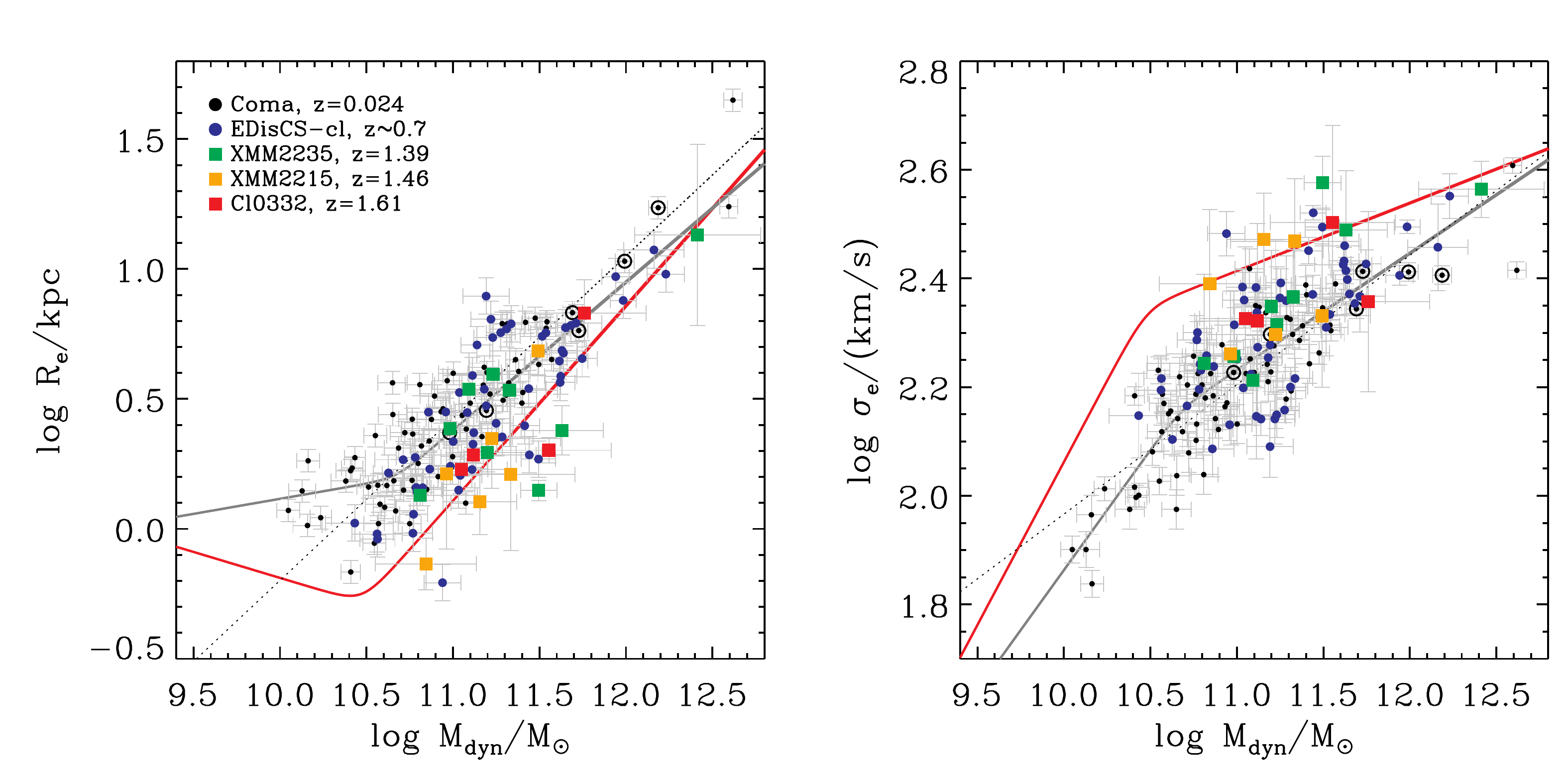} 
\includegraphics[width=0.8\textwidth]{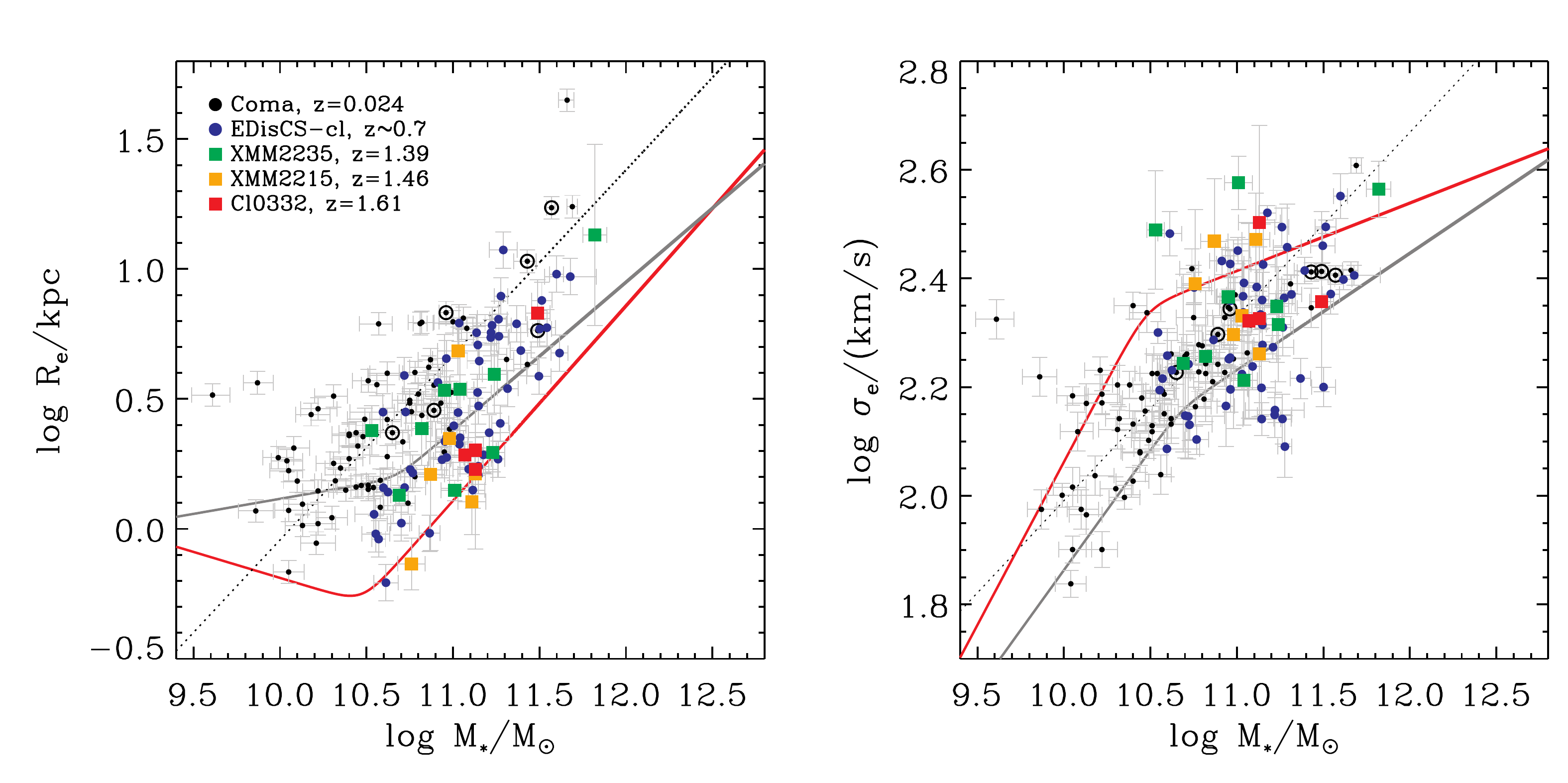}
\caption[] {{\em Left panels:} Size $-$ mass relation.  {\em Right
    panels:} Stellar velocity dispersion $-$ mass relation. {\em Upper
    panels:} Dynamical masses are used.  {\em Lower panels:} Stellar
  masses are used.  Symbols as in Figure~\ref{fig:fp_ml}.  Linear fits
  to the local data of Coma are shown with the dotted black lines; for
  comparison we overplot the \re$- M$ and \sigmae$- M$ relations from
  Equation~5 of \citet{Cappellari2013b} with the gray solid line. In
  red we show the zone of exclusion for local galaxies from Equation~4
  of \citet{Cappellari2013b}, \citet{Cappellari2016}.}
\label{fig:str_par}
\end{figure*}
%************************************************************************

\subsection[]{Effects of structural evolution on the FP zero-point evolution}
\label{subsec:FP_m_l_str}

\subsubsection[]{Zero point evolution and luminosity evolution}
\label{subsubsec:zero_lum_ev}

Following \citet{Saglia2010,Saglia2016} we can write the 
evolution of the FP zero-point in a generalized form including the terms related to the
structural and stellar velocity dispersion evolution as well as the
term describing the variation of the luminosity with
redshift. Thus, the luminosity variation can be written as:

\begin{equation}
\Delta \log L_{\rm FP, str. ev.}= \frac{2b+1}{b}\Delta \log R_{\rm e} - \frac{a}{b} \Delta \log\sigma_{\rm e}-\frac{1}{b}\Delta c_{z} 
\label{eq:str}
\end{equation}

\noindent
where $a$ and $b$ are the FP coefficients from Eq.~\ref{eq:fp}. We
parametrize each term of Eq.~\ref{eq:str} as a function of $\log(1+z)$
such that $\Delta \log R_{\rm e}=\nu\log(1+z)$, $\Delta \log
\sigma_{\rm e}=\mu\log(1+z)$ and $\Delta c_{z}=\eta' \log(1+z)$, where
$\nu$, $\mu$ are the slopes of the size and stellar velocity
dispersion evolutions with redshift, whereas $\eta'$ is related to the slope we
derive from the $\Delta \log M/L$ evolution with redshift; from Eq.~\ref{eq:m_l} $\eta'=\eta \times b$.

Eq.~\ref{eq:str} can therefore be written:

\begin{eqnarray}
\Delta \log L_{\rm FP, str. ev.}& =& \left(\frac{2b+1}{b} \nu-
                                     \frac{a}{b}  \mu -
                                     \frac{1}{b}\eta' \right)\log(1+z) \nonumber \\ 
                                          & = & \chi \log (1+z),
\label{eq:str2}
\end{eqnarray}

\noindent
where  $\chi=(\frac{2b+1}{b} \nu- \frac{a}{b} \mu -\frac{1}{b}\eta')$.

%************************************************************************
%          TABLE 3
%************************************************************************
\begin{table*}
\begin{scriptsize}
\begin{center}
\caption{Luminosity evolution as derived from the FP {  zero-point} with and without
  structural evolution, and from the luminosity-mass relation.}
\begin{tabular}{c c c c c c c c}
\hline
\hline
\noalign{\smallskip}
Case & Relation                                             &  \multicolumn{2}{c}{\xmm}                        &    \multicolumn{2}{c}{\xcs}               &   \multicolumn{2}{c}{\cl}          \\
     &                                                &  \mstar                   &        \mvir             & \mstar                 &        \mvir        &    \mstar      &        \mvir      \\                  
\noalign{\smallskip}
\hline
\noalign{\smallskip}
1 & $\Delta \log L_{\rm FP}= -\frac{1}{b}\eta' \log(1+z)$          &           \multicolumn{2}{c}{$1.68\pm0.37$}        & \multicolumn{2}{c}{$1.91\pm0.25$}          &  \multicolumn{2}{c}{$2.10\pm0.37$}  \\
2 & $\Delta \log L_{\rm FP, str. ev.}= \chi \log(1+z)$, $^{a}$        &  $1.01\pm0.41$       &   $1.56\pm0.48$             &  $1.25\pm0.31$         & $1.79\pm0.39$    &   $1.44\pm0.42$  & $1.99\pm0.48$  \\
3 & $\Delta\log L_{\rm L-mass}= \tau \log(1+z)$, $^{a}$             &  $1.44\pm0.12$        &  $1.72\pm0.26$              & $1.18\pm0.22$         & $1.97\pm0.21$     &   $1.07\pm0.14$ &  $2.04\pm0.26$  \\
\hline
\label{tab:cluster_fit_lumi}
\end{tabular}

\begin{minipage}{17cm}
  {\sc Notes.} --- The evolution of the galaxy luminosity as derived
  from the FP $L_{\rm FP}$, luminosity evolution including 
  both structural evolution and FP zero-point evolution as described
  in Eq.~\ref{eq:str2} $L_{\rm FP, str. ev.}$, and luminosity
  evolution from the mass-luminosity relation $L_{\rm
    L-mass}$. $^{a}$To allow enough dynamic range to fit the
  size-mass, stellar velocity dispersion-mass and luminosity-mass
  relation, we adopt the full sample with
  $\log M_{\star}/M_{\odot}>10.5$.  Results are consistent within the
  errors if we trace the evolution of the luminosity only for the
  $\log M_{\star}/M_{\odot}>11$ sample.
\end{minipage}

\end{center}
\end{scriptsize}
\end{table*}
%************************************************************************

We first assume that the FP evolution is only due to the $M/L$
evolution with redshift as a result of aging stellar population. This
means that the two terms $\Delta \log R_{\rm e}$,
$\Delta \log \sigma_{\rm e}$ in  Eq.~\ref{eq:str} are zero and
$\Delta \log L_{\rm FP}=-\frac{1}{b}\eta' \log (1+z)=-\eta \log (1+z)$.
Table~\ref{tab:cluster_fit_lumi}, case 1) shows the results of the fit.

We then include in the luminosity evolution the effect of varying
structural properties with redshift in the evolution of the
fundamental plane; this means including all terms of
Eq.~\ref{eq:str2}, and propagating the errors consistently.

\subsubsection[]{Size-mass and stellar velocity dispersion-mass
relations}
\label{subsubsec:size_sigma_ev}

We study how the size-mass and stellar velocity dispersion-mass
relations vary with redshift for our "dispersion'' sample adopting the
Coma data as local reference, and by considering both stellar masses
and dynamical masses (see Figure~\ref{fig:str_par}).
For both relations we assume that the slope does not change with
redshift and adopt the value we obtain from our Coma sample as done in
previous work (e.g., \citealt{Saglia2010}, \citealt{Newman2012}).
In \citet{Chan2016} and Chan et al sub. we present a more detailed
analysis of the size-mass relation for the full red-sequence sample of
the three clusters and confirm this assumption.

We note that the \re$-$\mstar\ and \sigmae$-$\mstar\ relations are
actually part of a trend of size and stellar velocity dispersion with
age and morphology (see Figure~1 of \citealt{Cappellari2013b} and
Figures~20-23 of \citealt{Cappellari2016}) and, if this was not
accounted, different sample selections could be mistaken for
structural evolution.
For this reason we also include in Figure~\ref{fig:str_par} the zone
of exclusion for local galaxies from \citet{Cappellari2013b},
\citet{Cappellari2016}; this region corresponds to a lower limit for
the existence of local passive galaxies in the diagrams. For the left
panels of Figure~\ref{fig:str_par} we derive the zone of exclusion
using Equation~4 of \citet{Cappellari2013b}, and rescale the size
along the semi-major axis given by that equation, to a circularized
size - as used in this work $-$ adopting the median axis ratio of our
Coma sample ($\sim 0.65$). For the right plots of
Figure~\ref{fig:str_par} we convert the zone of exclusion using the
virial relation $M=5.0\times\sigma_{\rm e}^2 R_{\rm e} /G$ following
the prescriptions of \citet{Cappellari2013b},
\citet{Cappellari2016}. We also overplot the \re$-M$ and \sigmae$-M$
relations from Equation~5 of \citet{Cappellari2013b} for comparison
with our sample.

We note that the \re$-$\mstar\ and \sigmae$-$\mstar\ we find from our
sample when adopting dynamical masses (black dotted lines in the upper
panels of Figure~\ref{fig:str_par}) are consistent with those of
\citet{Cappellari2013b} (gray solid lines in Figure~\ref{fig:str_par})
in the range of \mvir$>$10.67. In this case we also see that the
sample of galaxies with the oldest ages in Coma are closer to the zone
of exclusion as expected based on the results of
\citet{Cappellari2016}; about half of our KCS sample is either
below/above the zone of exclusion defined by the \re$-M$ and
\sigmae$-M$ trends, respectively, as expected in the case of
significant size evolution.  If we adopt stellar masses we see a
zero-point offset in our fitted relation compared to that of
\citet{Cappellari2013b}, probably due to a $\sim$0.3 dex offset
between dynamical and stellar masses in the Coma sample (see also
Section~\ref{sec:local}). This offset could be due to effects of
non-homology, change of the IMF or dark matter fractions;
understanding this offset is beyond the scope of this paper. For this
reason we test local scaling relations using both dynamical and
stellar masses in our work.

To study the evolution of the scaling relations with redshift we use
the approach followed by \citet{Newman2012}, \citet{Cimatti2012},
\citet{Delaye2014}, \citet{vanderWel2014} and Chan et al sub., where
we remove the correlation between \re\ and \mstar\ (or \mvir) or \sigmae\
and \mstar\ (or \mvir) by dividing sizes and stellar velocity dispersions
by a reference mass of $M\sim10^{11}$\msun. We then trace the
resulting quantities, which we call mass-normalized size
$\log R_{\rm e, mass-norm}$ and stellar velocity dispersion
$\log \sigma_{\rm e, mass-norm}$, as a function of redshift.  This
step is necessary when comparing samples with different mass
distributions.
Once the slope of the size-mass and stellar velocity dispersion-mass
relation is assumed (see above), this procedure is equivalent to
tracing the evolution of the zero-point of the \re\ and \mstar\ or
\sigmae\ and \mstar\ relations.

We trace the variation as a function of redshift of
$\Delta \log R_{\rm e, mass-norm} \propto \nu \log(1+z)$ and
$\Delta \log \sigma_{\rm e, mass-norm} \propto \mu \log(1+z)$ and
derive the slopes $\nu$ and $\mu$.
The results for both stellar mass and dynamical-mass normalized
quantities are shown in Table~\ref{tab:cluster_fit_size_ev}. We fit
the three KCS clusters together because the variation with redshift of
\re\ and \sigmae\ have similar dynamic range for the three clusters,
moreover we include the full sample with $\log
M_{\star}/M_{\odot}>10.5 $ in the fit.
We note that \citet{vanderWel2014} evolution of the median sizes in
our mass bin in the field is slightly steeper than what we find in
this work possibly related to differences between cluster and field
sample {  (see also Chan et al sub.)}.

Median $R_{\rm e, mass-norm}$ for the three clusters are 55\%
smaller than median $R_{\rm e, mass-norm}$ of Coma galaxies when
stellar-mass normalized sizes are used, and 38\% smaller when
dynamical-mass normalized radii are used.
The median $\sigma_{\rm e, mass-norm}$ in the KCS sample is 3\%
larger than the median $\sigma_{\rm e, mass-norm}$ of Coma when
stellar-mass normalized quantities are used, and 20\% larger, when using
dynamical-mass normalized stellar velocity dispersions.
This does not change if we use the subsample of Coma galaxies whose
age is $> 9$ Gyr: there is a $5-7$\% difference between the median
$R_{\rm e}$ of the whole Coma sample and the subsample with an age $> 9$ Gyr;
the difference reaches up to 20\% for the median stellar velocity
dispersions.

KCS galaxies follow similar structural and stellar velocity dispersion
evolution as the EDisCS-cluster sample \citep{Saglia2010}.  Our
results for the size evolution are in partial conflict with recent
work by \citet{Jorgensen2014} where almost no size variation with
redshift is found from their sample of clusters at $z<1$ \cite[from
e.g.,][]{Jorgensen2013}; their cluster at $z>1.27$ shows trends
similar to ours, suggesting that larger effects of size-evolution can
be seen at $z>1$. We note that the tension with the cluster sample at
$z<1$ could also be related to the different selections used in our
and \citet{Jorgensen2014} samples.

\subsubsection[]{Effects of structural evolution on derived ages}
\label{subsubsec:effects_zp}

The net contribution of structural and stellar velocity dispersion
evolution $\chi_{\rm str-ev} =(\frac{2b+1}{b} \nu- \frac{a}{b} \mu)$
is {  $-0.67$} and $-0.12$ (a mean {  $\sim$35\%} and {  6\%} of
the FP zero-point evolution for our sample), when we consider
relations of stellar mass or dynamical mass normalized, respectively.
$\chi_{\rm str-ev}$ is smaller than the $\eta$ in case 1) of
Table~\ref{tab:cluster_fit_lumi} suggesting that most of the
zero-point evolution is indeed driven by the evolution of the
luminosity.  As mentioned above, the structural evolution we derive
with stellar mass-normalized quantities is larger, which also affects
the slopes in case 2) of Table~\ref{tab:cluster_fit_lumi} where we
find shallower slopes.

{  We test what is the effect on the derived ages by rescaling up
  the $\Delta \log M/L$ of each overdensity of an amount corresponding
  to the percentage difference between the slopes of case 1) and case
  2) of Table~\ref{tab:cluster_fit_lumi}. By using mass-normalized
  slopes of case 2) we find that the mean age of the $\log
  M_{\star}/M_{\odot}>11$ galaxies in \xmm\ becomes larger than the
  age of the universe, which is unfeasible. We therefore consider as
  upper limit the age of the universe, which is a factor $\sim 2$ than
  the age we derive without considering structural evolution. For
  \xcs, we find that the mean age becomes a factor $\sim 2.4$ larger,
  and for \cl\ we find as well a mean age a factor $\sim2.4$
  larger than that we obtained when we do not include structural
  evolution in our analysis (case 1)). The large ages we find suggest
  that the structural evolution we estimate by using stellar-mass
  normalized \re\ and \sigmae\ could be overestimated. This could be
  due to a stronger ``progenitor bias'' when selecting galaxies in
  \mstar, for instance.  The sample of Coma galaxies for which we
  could compare the \re\ and \sigmae\ distributions for the oldest
  population is limited (i.e., only 6 galaxies have an age $>9$ Gyr);
  a larger sample of ages for Coma galaxies would be helpful to solve this
  issue.

  By adopting dynamical mass-normalized slopes of case 2) we find
  larger ages $-$ though consistent within the errors $-$ compared to
  those resulting from case 1).  For \xmm\ the mean age becomes $\sim
  0.44$ Gyr larger, for \xcs\ $\sim 0.22$ Gyr larger and for \cl\
  $\sim 0.18$ Gyr larger than the case in which structural evolution
  is not accounted.}

%************************************************************************
%          TABLE 4
%************************************************************************
\begin{table}
\begin{scriptsize}
\begin{center}
\caption{Redshift evolution of \re\ and \sigmae.}
\begin{tabular}{c c c}
\hline
\hline
\noalign{\smallskip}
                     &       \mstar -normalized              &        \mvir   -normalized                    \\
\hline
Relation &    slope       &        slope             \\
\noalign{\smallskip}
\hline
\noalign{\smallskip}
$\Delta \log R_{\rm e, mass-norm}\propto \nu \log(1+z)$  & $-1.04\pm 0.12$  & $-0.85\pm0.30$  \\
$\Delta \log \sigma_{\rm e, mass-norm}\propto \mu \log(1+z)$  & $0.09\pm0.10$   & $0.34\pm0.12$    \\
\hline

\label{tab:cluster_fit_size_ev}
\end{tabular}
\begin{minipage}{8cm}

  {\sc Notes.} --- Uncertainties on each parameter are 1$\sigma$
 {  Jackknife} errors.  The evolution of \re\ and \sigmae\ is calculated for the
  three clusters simultaneously as $\Delta \log R_{\rm e}\propto \nu
  \log(1+z)$, and $\Delta \log \sigma_{\rm e}\propto \mu \log(1+z)$,
  respectively.

\end{minipage}

\end{center}
\end{scriptsize}
\end{table}
%************************************************************************

\subsubsection[]{Luminosity evolution from luminosity-mass relation}
\label{subsubsec:lum_ev}

As an additional test, we compare the luminosity evolution derived
from {  fitting the luminosity-mass} relation as a function of
redshift.  For this we assume a constant slope of the luminosity-mass
relation as derived fitting the Coma sample. We follow a procedure
similar to the size and stellar velocity dispersion, and derive both
stellar mass and dynamical mass-normalized luminosities, adopting
$M\sim10^{11}$\msun\ as reference mass.  We note that while fitting
the luminosity-mass relation we include the full sample with $\log
M_{\star}/M_{\odot}>10.5$ in the fit to allow enough dynamic range. As
expected, the results do not change using only galaxies with $\log
M_{\star}/M_{\odot}>11$ because we normalize the luminosity.  This
test is described by case 3) of Table~\ref{tab:cluster_fit_lumi} where
$\Delta \log L_{\rm \rm L-mass}= \tau \log (1+z)$. {  Error bars
  are estimated with a Jackknife technique.}
The results would not change if we use the sample of galaxies with age
$>9$ Gyr rather than the full sample of Coma galaxies as reference.
The difference of the median luminosity of the full sample and that
with ages $>9$ Gyr, is 19\% for stellar-mass normalized luminosities
and 1\% for dynamical-mass normalized luminosities.
 
The three scenarios show consistent results when adopting
dynamical-mass normalized quantities, with in general steeper
luminosity evolution at constant dynamical mass, confirming the
limited impact of structural evolution in our sample (see also
\citealt{Saglia2010}, \citealt{Saglia2016}).  We note that the use of
stellar-mass normalized quantities results in shallower slopes that
{  (see slope for \cl\ in case 3) of
  Table~\ref{tab:cluster_fit_lumi}} are inconsistent with the results
from the FP zero-point.

\subsection[]{Cluster versus field comparison}
\label{subsec:cluster_vs_filed}

{  The current findings show that massive $\log
  M_{\star}/M_{\odot}>11$ galaxies in the three KCS overdensities have
  overall formation ages consistent within the errors.  There
  is a possible weak suggestion that galaxies in the massive and virialized
  cluster \xmm\ are older compared to the massive galaxies in the
  lower-density structure \cl, after accounting for the difference in
  the age of the Universe at the redshifts of the two overdensities.}
Similar {  results} are found by fitting stellar population models to the
stacked KMOS spectra (e.g., Houghton et al., in prep).

{  \citet{Mendel2015} found an age of $1.03_{-0.08}^{+0.13}$ Gyr
  from the analysis of stacked spectra of a sample of passive galaxies
  at the median redshift of $\langle z \rangle=1.75$, part of the KMOS
  VIRIAL field survey.  The ``redder'' part of their sample at
  $\langle z \rangle=1.73$ shows ages of $1.22^{+0.56}_{-0.19}$ Gyr,
  whereas the ``bluer'' part of their sample at $\langle z \rangle=1.82$
  is as young as $0.85^{+0.08}_{-0.05}$ Gyr.  We compare our KCS
  sample with the field sample of VIRIAL, by accounting for the
  difference between the redshift of the two samples. \footnote{We
    note that \citet{Mendel2015} used \citet{Conroy2009} SSPs to
    derive their ages from stacked spectra. In
    Appendix~\ref{subsec:ssp_Z} we show that if we were to use
    different SSPs in our FP analysis, ages would change at most of
    $\sim0.15$ Gyr.} Galaxies in \cl\ show an average age that is
  consistent with the average age of massive galaxies in the full
  VIRIAL sample, which is expected because \cl\ is by far the lowest
  density environment in our survey.  For \xmm\ we find consistent
  mean ages with the ``redder'' VIRIAL population. \xmm\ ages are also
  consistent with the ages of the ``blue'' VIRIAL population, but with
  lower significance.}
This could originate from an accelerated evolution of
galaxies in the most massive overdensities compared to field galaxies as
found in some FP studies at lower redshifts. 

\citet{Saglia2010} found a difference of about $\sim 1$ Gyr from the
$\Delta \log M/L_{B}$ evolution of cluster and field galaxies from the
EDisCS survey at $z\sim 0.7$, whereas \citet{Gebhardt2003} find $\sim
2$ Gyr difference for a sample of cluster and field galaxies at
$z\sim0.8$.  By comparing $\Delta \log M/L_{B}$ of cluster and field
galaxies up to $z\sim1.27$ \citet{vanDokkum2007} found an age difference
of $\sim 0.4\pm0.2$ Gyr, {  which is more in line with our findings}.
One of the possible sources of discrepancy {  between different
  studies} could be the amount of progenitor bias in the field sample
at $z>1$ \citep[e.g.,][]{Mendel2015}, which could affect the cluster
and field comparison \citep[e.g.][]{vanDokkum2007}.
Our findings {  could be} consistent with a scenario in which there is a link
between the time of assembly of a cluster and the stellar population
of the galaxies residing in it. Semi-analytical models for instance
\citet{DeLucia2006} show an age difference of $\sim 0.7$ Gyr between
massive galaxies in overdense and underdense regions of the Universe.

By also including galaxies down to $\log M_{\star}/M_{\odot}=10.5$ in
the \xmm\ sample, we find younger ages, supporting previous findings
of trends of age with \mstar\
\citep[e.g.,][]{Treu2005,Renzini2006,vanderMarel2007}.  For \xcs\ the
age does not change appreciably, which is probably related to some
bias in the sample for which we derived stellar velocity dispersions.

%************************************************************************
% FIGURE 8
%************************************************************************
\begin{figure}
\centering
\includegraphics[width=\columnwidth]{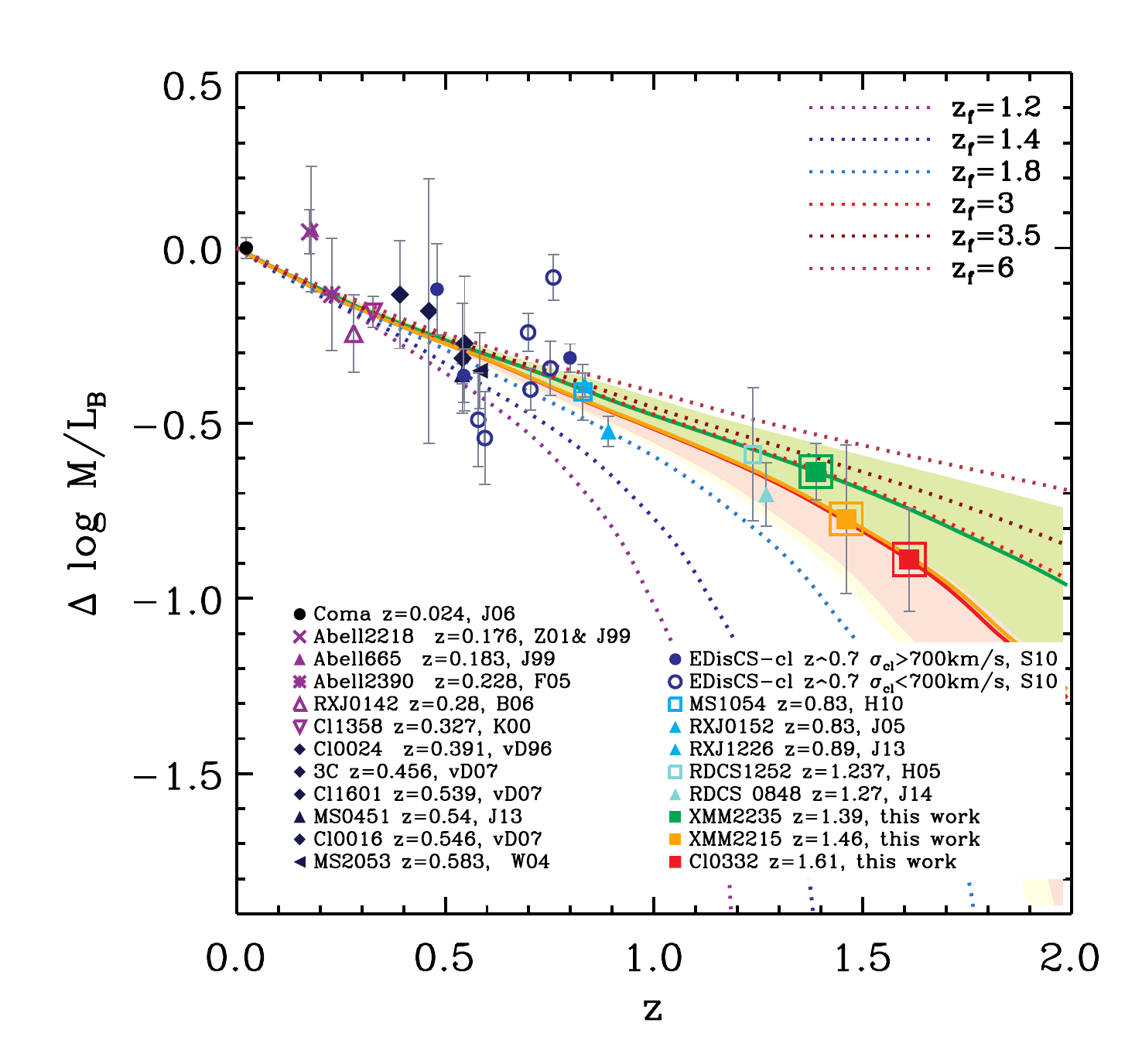}
\caption[] {Redshift evolution of the {  weighted-mean} $\Delta \log M/L_{B}$ for
  the $\log
M_{\star}/M_{\odot}>11$  galaxies in each KCS cluster and for the $\log
M_{\star}/M_{\odot}>11$ galaxies in clusters in
  the literature. Symbols and lines as in Figure~\ref{fig:summary_FP1} for the
  sample described in this paper.  The literature sample is as follows:
  Coma, black filled circle \citep{Jorgensen2006}; Abell 2218, purple
  cross, was derived averaging the values from \citet{Jorgensen1999b}
  and \citet{Ziegler2001}; Abell 665, purple filled triangle
  \citep{Jorgensen1999b}; Abell 2390, purple star
  \citep{Fritz2005}, RXJ0142, purple open triangle
  \citep{Barr2005,Barr2006}; Cl1358+62, purple open downward triangle
  \citet{Kelson2000}; Cl0024+16, navy filled diamond
  \citep{vanDokkum1996}; 3C295 Cl1601+42 and Cl0016+16, navy filled
  diamonds \citep{vanDokkum2007}; MS 0451.6-0305 navy filled triangle
  \citep{Jorgensen2013}; MS 2053-04 navy filled left-facing triangle
  \citep{Wuyts2004}; ``EDisCS-cl'' with a $\sigma_{\rm cl}>700$\kms,
  blue filled circles \citep{Saglia2010}; `EDisCS-cl'' with a
  $\sigma_{\rm cl}<700$\kms, blue open circles \citep{Saglia2010}; MS
  1054-03, cyan empty square \citep{Holden2010}; RXJ0152-13, cyan
  filled triangle \citep{Jorgensen2005}; RXJ1226+33, cyan filled
  triangle \citep{Jorgensen2013}; RDCS1252.9-29, turquoise open square
  \citep{Holden2005}, RX J0848.6+44, turquoise filled triangle
  \citep{Jorgensen2014}.}
\label{fig:summary_FP3}
\end{figure}
%************************************************************************

\subsection[]{Comparison with the literature}
\label{subsec:comp_lit}

Our results for \xmm\ are consistent with \citet{Rosati2009}, who
inferred the ages of passive galaxies in the core and in the outskirts
of this cluster by analyzing spectro-photometric data available from
VLT/FORS2 and \hst. They found that massive galaxies in the core of
the cluster were formed at a formation redshift $z_{\rm f} > 3-4$ (we
find a formation redshift of $\sim 3$ for galaxies with
$\log M_{\star}/M_{\odot}>11$).  Similarly, we agree with
\citet{Lidman2008} and \citet{Strazzullo2010}, who found formation
redshifts $z_{\rm f} > 3$ using both mean colors and scatter of the
red sequence and luminosity function, respectively.  {  We note also
  our sample shows two main concentrations in the
  $\Delta \log M/L_{B} - z$ plane suggesting that there is probably
  an "older" and a "younger" population in the cluster.}

Previous work on \xcs\ show that galaxies in the cluster core have
some level of star formation
\citep[e.g.][]{Hilton2009,Hilton2010,Hayashi2010,Hayashi2011,Ma2015,Hayashi2017,
  Stach2017}.
In our work we find that galaxies are on average slightly younger than
the galaxies in \xmm\ (but still consistent within the errors), which
could fit in a scenario in which for this cluster star
formation is still happening while galaxies are falling into the
denser cluster environment. However, we caution this interpretation
given that for this cluster we derived stellar velocity dispersion
mostly in objects which are in the bluer part of the red sequence.
The mean age for the galaxies in \xcs\ is consistent with the lower
limit on the formation redshifts for the galaxies in the red sequence
of the cluster found by \citet{Hilton2009} using the scatter and
intercept of the color-magnitude diagram with respect to Coma.

The relatively young mean age we derived for the galaxies in \cl\ are
in agreement with the ages found by \citet{Cimatti2008} and
\citet{Kurk2009}.

In Figure~\ref{fig:summary_FP3} we compare the {  weighted-}mean
$\Delta \log M/L$
evolution of the KCS sample with $\Delta \log M/L$ available in the
literature for $\log M_{\star}/M_{\odot}>11$ galaxies at
$0.024<z<1.27$ in clusters with a wide range of mass and virialization
status (see caption of Figure~\ref{fig:summary_FP3} for details).  We
homogenize photometric data from different samples to a common
$B$-band surface brightness within the effective radius, following
procedures similar to Appendix A of \citet{vanDokkum2007}.  {  Error
  bars are uncertainties on the weighted mean.}
We find that our clusters extend up to $z=1.61$ the trends we see at
intermediate redshift, and expand the statistic in a redshift range
currently almost unexplored.  There are some weak hints that the most
massive clusters have older formation redshifts, though errors are
large and we cannot constrain {  those statements}. We note for
instance that our results for the $\log M_{\star}/M_{\odot}>11$
galaxies in \xmm\ have formation redshifts of the order of the
galaxies in the massive cluster RDCS1252.9-2927 at $z\sim 1.237$
\citep{Holden2005}. At intermediate redshift we expect the differences
to be more difficult to detect, and we indeed see a range of formation
redshifts.

A larger number of clusters with different mass and properties at { 
  the same redshift we studied in this paper as well as at} higher
redshift, will provide additional constraints to the scenario we
described above.

\section{Conclusions}
\label{sec:conclusions}

In this paper we present new results on the evolution of the FP of a
sample of 19 passive galaxies in dense environments at $1.39<z<1.61$ from
KCS, a GTO survey using KMOS at the VLT.  

Over the past 3 years, KCS observed $\ge$ 20 massive ($\log M_{\star}/M_{\odot}>10.5$)
passive galaxies in four main overdensities at $1.39<z<1.8$,
with a range of masses and
properties, as well as a lower-priority overdensity at $z=1.04$
to bridge our high-redshift observations with the local sample.
With KCS we systematically targeted a large sample of galaxies in the
red sequence and built a new sample of stellar velocity dispersions in
{\it dense environments} at $z>1.39$. 

In this paper we present the analysis of the KMOS data for the sample
at $1.39<z<1.61$. KMOS data are combined with the structural parameters
derived from \hst\ imaging for the same galaxies and we obtain the
formation age through the fundamental plane.

Our main results can be summarized as follows:

\begin{itemize}

\item The zero-point of the $B$-band FP evolves with redshift, such
  that the highest-redshift cluster has the largest offset from Coma.
  By converting the zero-point evolution into an evolution of $\Delta
  \log M/L_{B}$ with redshift we find that $\log
  M_{\star}/M_{\odot}>11$ galaxies have $\Delta \log
  M/L_{B}=(-0.46\pm0.10)z$ in \xmm\, to $\Delta \log
  M/L_{B}=(-0.52\pm0.07)z$ in \xcs, and $\Delta \log
  M/L_{B}=(-0.55\pm0.10)z$ for \cl, respectively.
  The $\Delta \log M/L_{B}$ becomes steeper when we include less massive
  ($10.5<\log M_{\star}/M_{\odot}<11$) objects suggesting a trend with
  mass of the $\Delta \log M/L_{B}$.

\item By using \citet{Maraston2005} single stellar population models
  we derived mean formation ages for the sample with $\log
  M_{\star}/M_{\odot}>11$ from the $\Delta \log M/L_{B}$ evolution
  with redshift. We find mean luminosity-weighted ages to be
  $2.33^{+0.86}_{-0.51}$ Gyr, $1.59^{+1.40}_{-0.62}$ Gyr and
  $1.20^{+1.03}_{-0.47}$ Gyr, for \xmm, \xcs, and \cl,
  respectively. If we include in the sample of \xmm\ objects with
  $10.5<\log M_{\star}/M_{\odot}<11$ we derive younger ages
  $\sim1.63^{+0.39}_{-0.29}$ Gyr, suggesting a trend of age with
  \mstar. For \xcs\ we obtain consistent ages $\sim
  1.62^{+1.30}_{-0.61}$ Gyr.

\item Our results are robust against the use of different SSPs or
  metallicity assumptions. Formation ages are also consistent with the
  expectation for the color evolution from SSPs.

\item Effects of structural and stellar velocity dispersion evolution
  are responsible for {  $\sim$6-35}\% of the evolution we see in
  the FP zero-point, most of which instead comes from the $M/L$
  evolution. {  The net impact of structural and stellar velocity
    dispersion evolution on the measurements of the galaxy ages is
    $\sim 2-2.4$ Gyr when we consider structural and velocity
    dispersion evolution normalized by the stellar mass (i.e., 35\%
    contribution in the FP zero-point evolution). Ages vary at most
    $\sim 0.44$ Gyr for \xmm\ when using structural and velocity
    dispersion evolution normalized by the dynamical mass; for \xcs\
    and \cl, the variation is smaller.} If we fit the luminosity-mass
  relation we find similar evolution of the luminosity with redshift
  to that we find in the FP.

\item The mean $M/L$ of the galaxies in the three overdensities
  relative to Coma are consistent with passive evolution with
  formation {  ages consistent within the errors} for the three
  clusters. {  However, there is a weak suggestion that more massive
    and virialized cluster are formed at earlier times compared with
    galaxies in lower-density structure in our sample.  This is
    consistent with our findings from the stellar population analysis
    of stacked KMOS spectra of the same galaxies as discussed in a
    companion paper (e.g., Houghton et al., in prep). We also find
    that massive $\log M_{\star}/M_{\odot}>11$ galaxies in \xmm\ have
    ages consistent with the "red'' population of passive galaxies in
    field at similar redshift from the VIRIAL KMOS GTO survey.  \xmm\
    ages are also consistent with the ages of the "blue'' VIRIAL
    population, but with lower significance.}

\end{itemize}

\acknowledgments

We thank the entire KMOS instrument and commissioning teams for their
hard work, which has allowed our observing program to be carried out
successfully.  We wish to thank the ESO staff, and in particular the
staff at Paranal Observatory, for their support during observing runs
over which the KMOS GTO observations were carried out.  
{  We acknowledge the anonymous referee for valuable comments that
  led to an improved presentation.}
We thank Matt Hilton for the reduced MOIRCS imaging, and Greg Rudnick
for providing the updated catalog of stellar masses for EDisCS
galaxies prior publication. AB thanks Paola Santini for providing the
CANDELS GOODS-S catalogs of stellar masses for comparisons prior to
appearance on the public website.  AB acknowledges Claudia Maraston
and Janine Pforr for the stellar mass catalog of SDSS/DR7 galaxies.
{  We would also like to thank Chris Lidman for providing HAWK-I
  images of XMMXCS J2215-1738 for the construction of our wide field
  of view photometry catalogs.}  JCCC acknowledges the support of the
Deutsche Zentrum f\"ur Luft- und Raumfahrt (DLR) via Project ID
50OR1513. DJW acknowledges the support of the Deutsche
Forschungsgemeinschaft via Project ID~3871/1-1 and ID~3871/1-2.  This
work was supported by the UK Science and Technology Facilities Council
through the `Astrophysics at Oxford’ grant ST/K00106X/1. RLD
acknowledges travel and computer grants from Christ Church, Oxford and
support from the Oxford Centre for Astrophysical Surveys which is
funded by the Hintze Family Charitable Foundation.  MC acknowledges
support from a Royal Society University Research Fellowship.  JPS
gratefully acknowledges support from a Hintze Research Fellowship.
This research has made use of the SVO Filter Profile Service
(http://svo2.cab.inta-csic.es/theory/fps/) supported from the Spanish
MINECO through grant AyA2014-55216.

%\bibliographystyle{apj}
%\bibliography{alebiblio}



\clearpage
\newpage

\appendix

\section{A.~~Derivation of the kinematics from different parts of the spectrum}
\label{sec:sigma_test}

%************************************************************************
% FIGURE APPENDIX A
%************************************************************************
\begin{figure*}
\centering
\includegraphics[width=\textwidth]{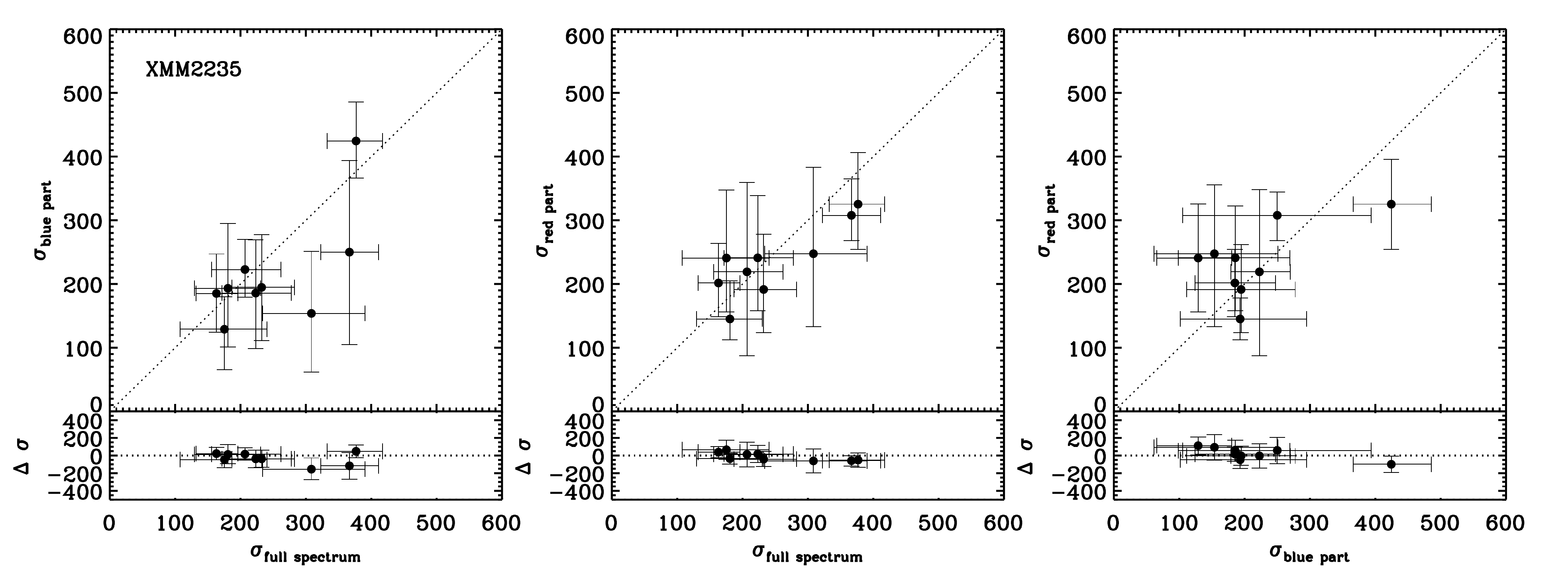}
\includegraphics[width=\textwidth]{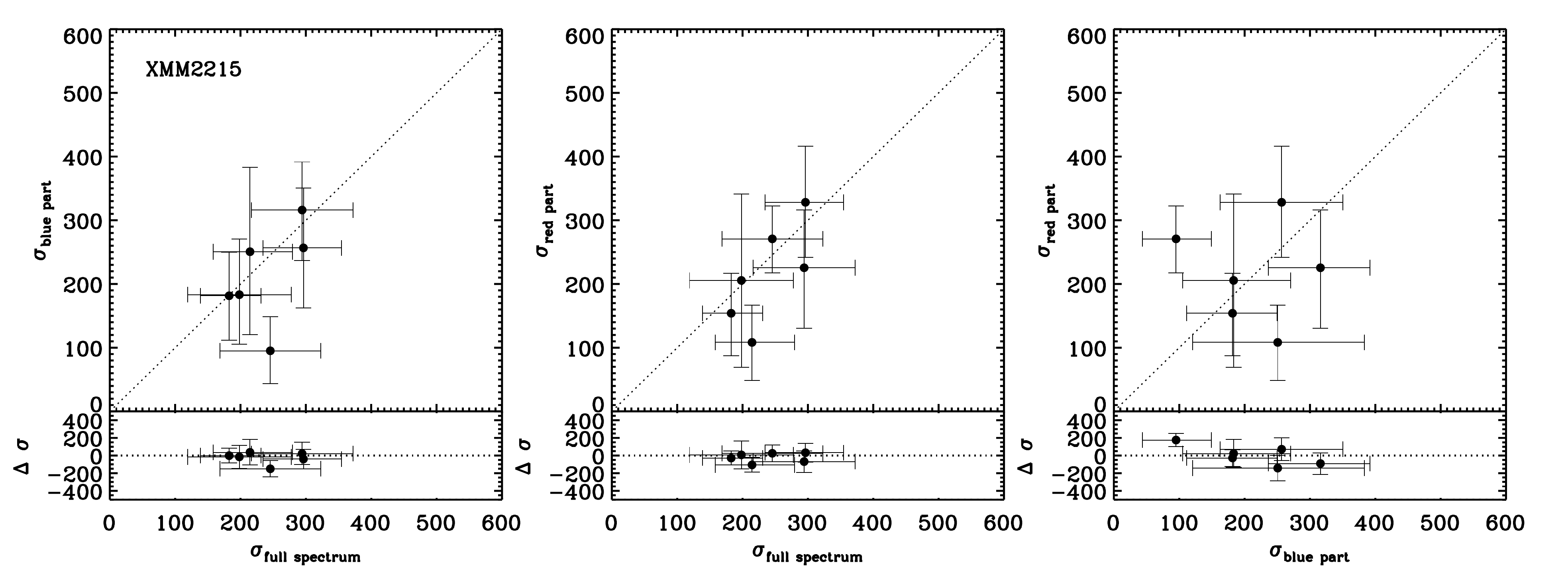}
\caption[] {Comparison between the stellar velocity dispersion
  measurements in the blue and red part of the spectrum for \xmm\
  (upper panel) and \xcs\ (lower panel). From left to right the panels
  show the comparison between the fit in the full spectrum and the
  blue part, the full spectrum and the red part, and between blue and
  red part, respectively.}
\label{fig:test_sigma2}
\end{figure*}
%************************************************************************

We assess the robustness of our fits by measuring the stellar velocity
dispersion separately in the blue and red parts of our spectra.  This
is done for the clusters \xmm\ and \xcs\, where multiple diagnostic
lines are available in regions free from telluric absorption or strong
sky emission. The blue part of the spectrum covers mostly the G-band
and H$_\gamma$, and also Ca line for \xcs, whereas the red part
includes mostly H$_\beta$ and $\rm Fe$ lines, and Mg for \xmm.
We note that the $S/N$ significantly decreases in the blue part of the
spectrum, due to the lower throughput of KMOS in the bluer part of the
$YJ$ band, affecting the uncertainties of our measurements.  Moreover,
the blue part of the spectrum we fit covers a short wavelength range
compared to the red part. This could potentially affect the set of
templates applied by the fitting procedure.  To circumvent this issue,
we adopt the same combination of templates derived in the fit of the
full spectrum in both the fit of the blue and red part.
As discussed in Section~\ref{subsec:sigma}, before performing the
kinematic fit we smooth the KMOS spectrum with a variable kernel to a
common FWHM to match the maximum FWHM.  This should prevent us to add
systematics related to the KMOS resolution in the blue and red part of
the spectrum.

Figure~\ref{fig:test_sigma2} shows a comparison of the results
obtained by fitting the full spectrum and the blue part (left panels),
the full spectrum and the red part (central panels) and red and blue
part (right panels). The actual value is determined as the mode of the
distribution over the 100 bootstrap realizations and the error as the
standard deviation.  The actual value of the stellar velocity
dispersion from our spectra is consistent within the errors with the
mode of the distribution of \sigmae\ derived from the bootstrap
realizations.

For \xmm\ we find that the kinematics derived from the blue part and
from the full spectrum have a median difference $\Delta \sigma
=-37.50$ \kms\ ($\Delta \log \sigma=-0.08$) and $1\sigma$ scatter of
$70.11$ \kms; whereas by fitting the red region of the spectrum we
find an offset $\Delta \sigma=-35.55$ \kms\ ($\Delta \log
\sigma=-0.06$) and $1\sigma$ scatter of $56.20$ \kms\ from the fit of
the full spectrum. The median difference between the fit of the blue
and red part of the spectrum is $\Delta \sigma = 16.41$ \kms\ ($\Delta
\log \sigma=0.04$) with $1\sigma$ scatter of $72.54$ \kms.
In \xmm\ the quality of some data in the ``blue'' part is particularly
poorer resulting in a few objects with systematically higher \sigmae\
compared to the full sample.

For \xcs\ we find that the kinematics derived from the blue part and
from the full spectrum have a median difference $\Delta \sigma=-7.99$
\kms\ ($\Delta \log \sigma=-0.02$) and $1\sigma$ scatter of $60.80$
\kms; whereas by fitting the red region of the spectrum we find an
offset of $\Delta \sigma=-10.53$ \kms\ ($\Delta \log \sigma=-0.03$)
and $1\sigma$ scatter of $60.00$ \kms\ from the fit of the full
spectrum. The median difference between the fit of the blue and red
part of the spectrum is $\Delta \sigma = -2.54$ \kms\ ($\Delta \log
\sigma=0.01$) with $1\sigma$ scatter of $121.41$ \kms.

{  This test aims at assessing the systematic effect we can have
  deriving stellar velocity dispersions from different absorption
  lines at different redshift. It did not result in a rejection of
  galaxies, which were mostly discarded at the stage of
  the fitting of the full spectrum through all the bootstrap
  realizations. Generally, the systematic offsets
  $\Delta \log \sigma_{\rm e}$ between the full spectrum fit and the two
  sub-regions of the spectra are usually $<10$\%, and smaller than the
  typical 10-40\% uncertainty we have by fitting the full
  spectrum. Single objects could potentially have larger offsets than
  the median value given above in one of the wavelength ranges, but
  this could be attributed to a poorer quality of the spectrum in that
  wavelength range.}

\section[]{B.~~Success rate and selection functions}
\label{sec:FP_m_l_only_selection}

In this Section we describe the technique used to assess the selection
effects for the three overdensities in our sample.
We derive selection weights by assigning a selection probability to
each galaxy with a method similar to that used by \citet{Saglia2010}.

We first derive the completeness in measuring stellar velocity
dispersion for each of the clusters in our sample.  We split the red
sequence of our CMDs in equally-spaced magnitude bins (see
Figure~\ref{fig:cmd}), and for each bin we compute the ratio of the
number of red-sequence galaxies with measured stellar velocity
dispersion to the total number of galaxies in the red sequence in that
bin.
For the three overdensities we use different combinations of
magnitudes in the CMDs, therefore we derive separate weights for the
three overdensities, based on the three different photometries. 
We assign a probability to each galaxy by linearly interpolating these
selection curves. As we could expect, the stellar velocity dispersion
completeness is larger at brighter magnitudes.

We then rescale this fraction by the ratio between the number of
objects found to be spectroscopically confirmed members of the cluster
over the number of galaxies we targeted with KMOS for each magnitude
bin in the color magnitude diagram; those represent our selection
weights $\rm P_{\rm s}$.

We note that for \xmm, where we targeted $83$\% of the objects in the
red sequence with $H_{\rm F160w}<22.3$ we find that half of the
objects at $21.7<H_{\rm F160w}<22.3$ are either background or
foreground objects, confirming our expectations from the statistical
analysis in Section~\ref{subsubsec:target_sel}. Our findings support
also the results of Chan et al sub., who reached the same
conclusions using two-color diagrams.

%************************************************************************
% FIGURE  APPENDIX B
%************************************************************************
\begin{figure*}
\centering
\includegraphics[width=0.33\textwidth]{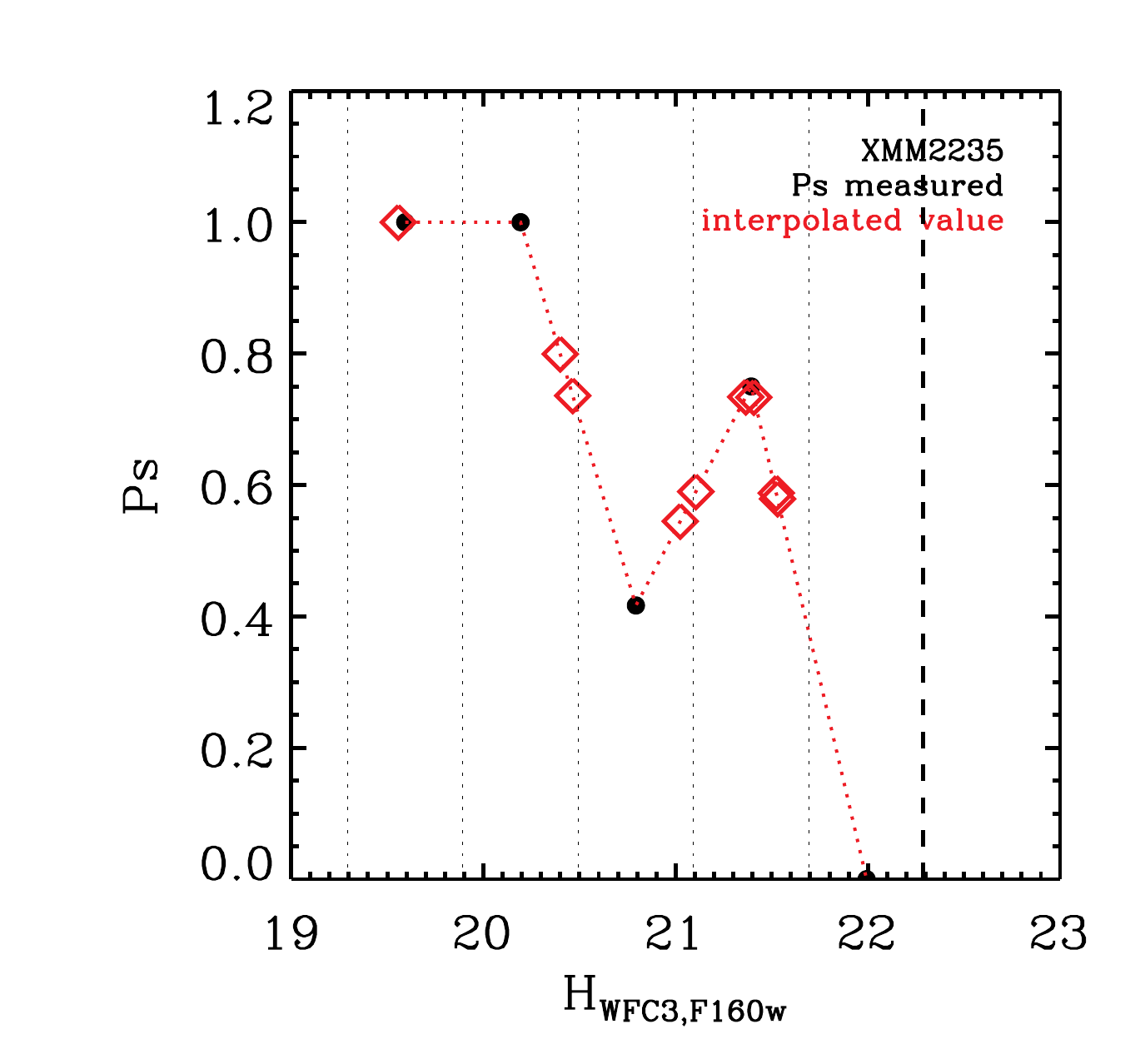}
\includegraphics[width=0.33\textwidth]{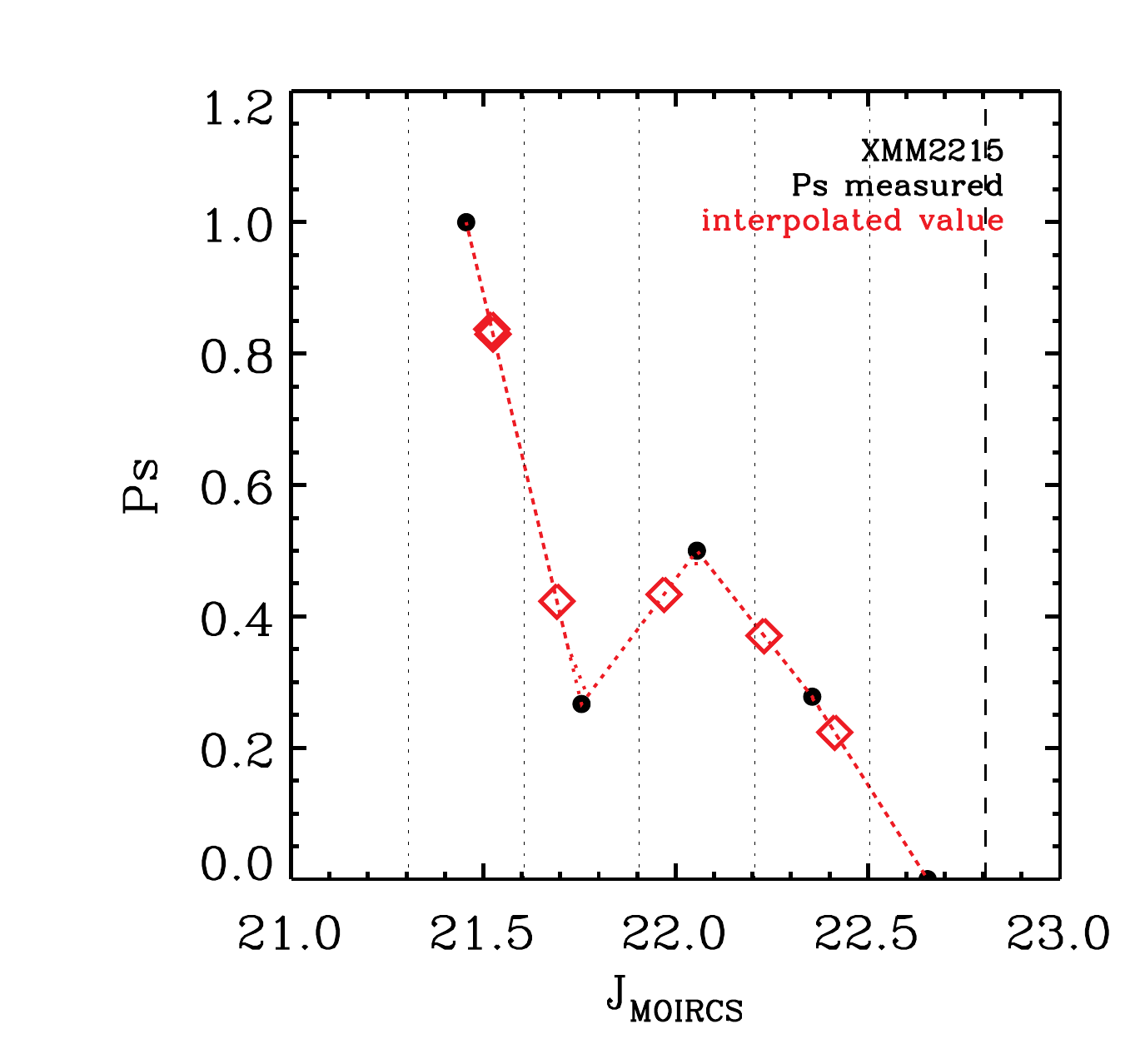}
\includegraphics[width=0.33\textwidth]{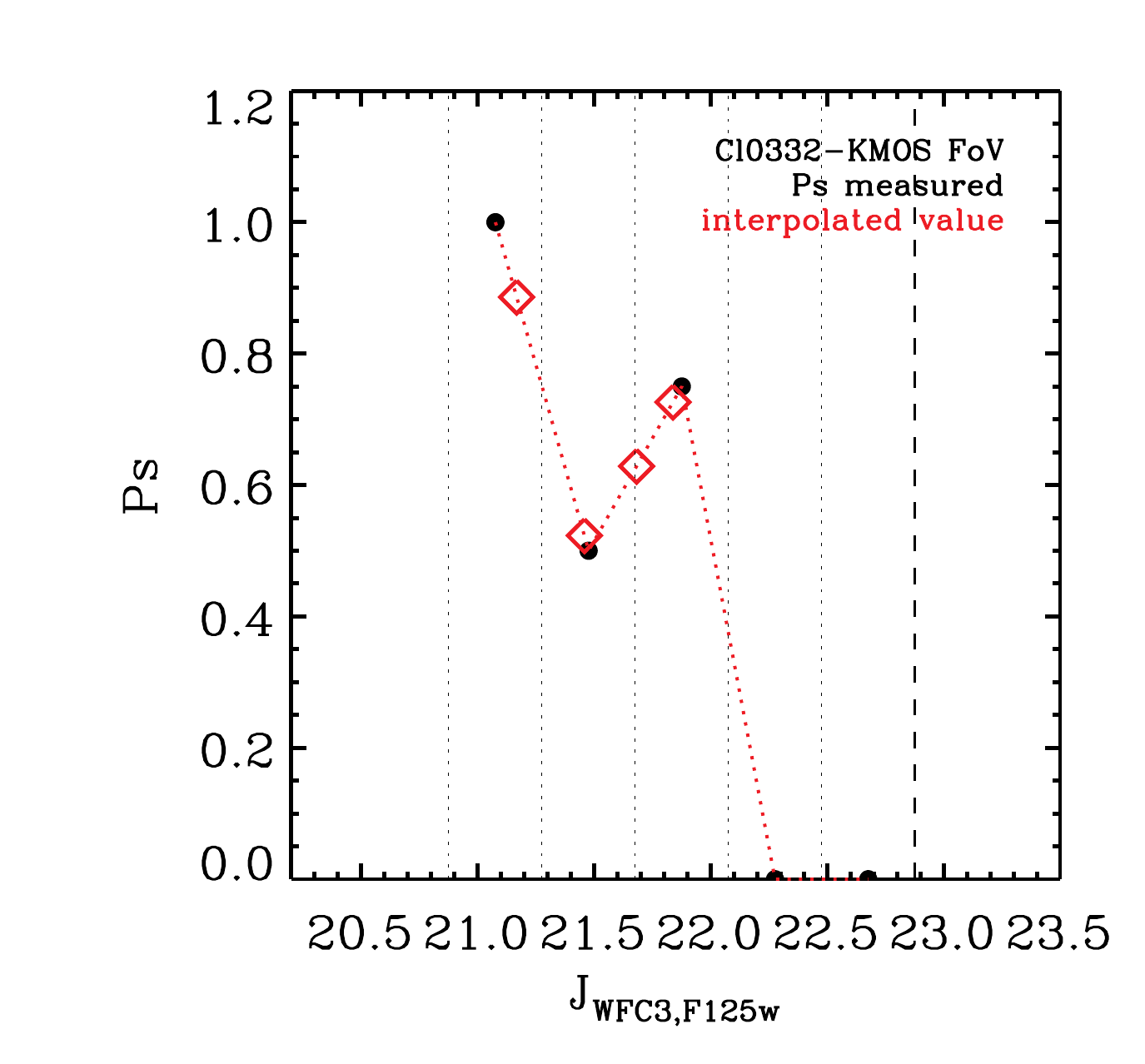}
\caption[] {The completeness
  functions $\rm P_{\rm S}$ of our cluster galaxies showing how many
  objects in the red sequence of our clusters have measured stellar
  velocity dispersions, normalized by how many objects were
  spectroscopically found to be members of the clusters over those
  allocated in the red sequence. Black circles show the average
  quantities per bin, and red diamonds the interpolated values
  corresponding to the magnitudes of the galaxies in our sample. Vertical
  dotted lines show the magnitude bins used to evaluate the selection
  functions for our sample (see also Figure~\ref{fig:cmd}). From
  left to right we show the functions for \xmm, \xcs\, and \cl,
  respectively.}
\label{fig:selection}
\end{figure*}
%************************************************************************

Figure~\ref{fig:selection} shows the weights as a function of the
selection magnitude, which is used for each cluster in
Figure~\ref{fig:cmd}.  The dashed lines show our limiting magnitude
for the KMOS observations of the three clusters and the dotted lines
show the magnitude bins as already shown in
Figure~\ref{fig:cmd}. Black dots are the average points used to derive
the curves, and the red lines and diamonds show the interpolated
values at the magnitude of each galaxy.
For \cl\ we derive weights for both the subsample of galaxies within
the region of the overdensity described by the field of view of our
KMOS pointing (see Figure~\ref{fig:selection}), as well as all the
objects within $\pm 3000$\kms\ of the overdensity redshift extending
over the whole GOODS-S field. In the latter case we find similar
weights for the two brightest bins, and lowers weights for the third
magnitude bin due to the larger number of objects in the red sequence.
The completeness in the velocity dispersion measurement for \cl\ is
higher at bright magnitudes and decreases towards fainter objects. The
trend is less clear for \xcs\ which we observed to a similar exposure
time.

We test whether the weights correlate with galaxy properties. \xmm\
does not show strong correlations, whereas for \xcs\ and \cl, weights
tend to be larger for more luminous and more massive objects.

\section[]{C.~~Effect of using different stellar population models and metallicity assumptions}
\label{subsec:ssp_Z}

\citet{vanDokkum2007} (their Figure~7) already showed that the
evolution of the $\Delta \log M/L_{B} $ for a SSP is similar in
\citet{Maraston2005} and \citet{Bruzual2003} SSP in the age range
between 9-10 Gyr in $B$ band. In this Section we assess whether this is still
valid in the redshift range of our overdensities.

We generate $M/L$ in the $B$ band as a function or redshift and
formation redshift using the {\tt EzGal} code of \citet{Mancone2012} for
both \citet{Bruzual2003} and \citet{Conroy2009}, \citet{Conroy2010}
SSPs with solar metallicity. We also produce $M/L$ for
\citet{Bruzual2003} SSPs with super-solar metallicities of
[Fe/H]$\sim0.56$ and sub-solar metallicity of [Fe/H]$\sim-0.33$.
As it is known, \citet{Bruzual2003} and \citet{Conroy2009,Conroy2010} SSPs
give similar $\Delta \log M/L_{B} $ with redshift, and give a
negligible difference in the fitted parameters, therefore in the
following we will quote only results based on \citet{Bruzual2003}.

We repeat the analysis described in Section~\ref{subsec:FP_m_l_only}
minimizing the difference between $\Delta \log M/L_{B} $ of the
\citet{Bruzual2003} SSPs with different formation redshifts (or age) and
the $\Delta \log M/L_{B} $ of our KCS sample with $\log
M_{\star}/M_{\odot}>11$.  
Ages differ at most of $0.15$ Gyr between using \citet{Maraston2005}
and \citet{Bruzual2003} SSPs {  in the $B$ band}.
As expected, this difference decreases with increasing
formation redshift, when the model give almost the same result.

We find consistent formation redshifts if we trace the color
evolution of galaxies with redshift using \citet{Bruzual2003} SSPs.

Our original assumption about the solar metallicity does not change
our results either within the errors. By adopting \citet{Bruzual2003}
SSPs with super-solar metallicities of [Fe/H]$\sim0.56$ (but assuming
solar metallicity at redshift 0 based on the known values of
metallicity for Coma) mean ages would become at most $\sim0.6$ Gyr
younger than with solar metallicity, whereas in case of sub-solar
metallicity of [Fe/H]$\sim-0.33$ (and assuming solar metallicity at
redshift 0) we derive ages that are at most $\sim1$ Gyr older.

In summary our results are robust against the use of different SSPs
and different metallicity assumptions and we maintain the same trends
we see using solar metallicity \citet{Maraston2005} SSPs.

\end{document}